\documentclass[useAMS,usenatbib]{mn2e}
\usepackage{hyperref}
\usepackage{graphicx}
 
\begin{document}

\title[ Detailed modelling of a large sample of \textit{Herschel} sources in Lockman]{Detailed modelling of a large sample 
of \textit{Herschel} sources in the Lockman Hole: identification of cold dust and of lensing
candidates through their anomalous SEDs
\thanks{{\it Herschel} is an ESA space observatory with science
instruments provided by European-led Principal Investigator
consortia and with important participation from NASA.}
}
\author[Rowan-Robinson M. et al]{Michael Rowan-Robinson$^{1}$, Lingyu Wang$^{2}$, Julie Wardlow$^{3}$, Duncan Farrah$^{4}$, 
\newauthor
Seb Oliver$^{5}$, Jamie Bock$^{6,7}$, Charlotte Clarke$^{5}$, David Clements$^{1}$, Edo Ibar$^{8}$, 
\newauthor
Eduardo Gonzalez-Solares$^{9}$, Lucia Marchetti$^{10}$, Douglas Scott$^{11}$, Anthony Smith$^{5}$, 
\newauthor
Mattia Vaccari$^{12}$, Ivan Valtchanov$^{13}$\\
${1}$Astrophysics Group, Imperial College London, Blackett Laboratory, Prince Consort Road, London SW7 2AZ, UK,\\
$^{2}$Department of Physics, Durham University, South Rd, Durham DH1 3LE, UK,\\
$^{3}$Dark Cosmology Centre, Niels Bohr Institute, University of Copenhagen, Denmark,\\
$^{4}$Dept of Physics, Virginia Polytechnic Institute and State University, 850 Wst Campus Drive, Blacksburg, VA 24061, USA,\\
$^{5}$Astronomy Centre, Dept. of Physics \& Astronomy, University of Sussex, Brighton BN1 9QH, UK,\\
$^{6}$California Institute of Technology, 1200 E. California Blvd., Pasadena, CA 91125, USA,\\
$^{7}$Jet Propulsion Laboratory, 4800 Oak Grove Drive, Pasadena, CA 91109, USA,\\
$^{8}$Instituto de F\'isica y Astronom\'ia, Universidad de Valpara\'iso, Avda. Gran Breta\~na
1111, Valpara\'iso, Chile,\\
$^{9}$Institute of Astronomy, Madingley Road, Cambridge CB3 0HA,\\
$^{10}$Department of Physical Sciences, Open University,Walton Hall, Milton Keynes MK7 6AA,\\
$^{11}$Department of Physics \& Astronomy, University of British Columbia, 6224 Agricultural Road, Vancouver, BC V6T~1Z1, Canada,\\
$^{12}$University of Western Cape, 7535 Belliville, Cape Town, South Africa,\\
$^{13}$Herschel Science Centre, European Space Astronomy Centre, Villanueva de la Ca\~nada,  28691 Madrid, Spain.\\
}

\maketitle
\begin{abstract}
We have studied in detail a sample of 967 SPIRE sources with 5-$\sigma$ detections at 350 and 500$\mu$m and associations
with Spitzer-SWIRE 24$\mu$m galaxies in the HerMES-Lockman survey area, fitting their mid- and far-infrared, and submillimetre, SEDs
in an automatic search with a set of six infrared templates.  For almost 300 galaxies we  have modelled their SEDs individually
to ensure the physicality of the fits.  We confirm the
need for the new cool and cold cirrus templates, and also of the young starburst template, introduced in earlier work.  We also 
identify 109 lensing candidates via their anomalous SEDs and provide a set of colour-redshift constraints
which allow lensing candidates to be identified from combined {\it Herschel} and {\it Spitzer} data. 
The picture that emerges of the submillimetre galaxy population is complex, comprising ultraluminous and hyperluminous starbursts,
lower luminosity galaxies dominated by interstellar dust emission, lensed galaxies and galaxies with surprisingly cold (10-13K) dust.
11$\%$ of 500$\mu$m selected sources are lensing candidates.  70$\%$ of the unlensed sources are ultraluminous infrared
galaxies and 26$\%$ are hyperluminous.  34$\%$ are dominated by optically thin interstellar dust ('cirrus') emission, but
most  of these are due to cooler dust than is characteristic of our Galaxy.   At the highest infrared luminosities we see SEDs
dominated by M82, Arp220 and young starburst types, in roughly equal proportions.

\end{abstract}


\section{Introduction}
The combination of {\it Herschel} (Pilbratt et al 2010) and {\it Spitzer} data provides us with the first 3-500$\mu$m 
spectral energy distributions of large samples of galaxies, for which we can accurately determine the masses of cold 
dust and search for very
young, heavily obscured starbursts.  The HerMES (Herschel Multi-tiered Extragalactic Survey) wide-area surveys 
(Oliver et al 2012) have been targeted on 
fields in which we already have excellent {\it Spitzer} data.

Over the past twenty years increasingly sophisticated radiative transfer models for different types of infrared galaxy
have been developed, for example for  starburst galaxies
(Rowan-Robinson \& Crawford 1989, Rowan-Robinson \& Efstathiou 1993, Silva et al 1998, Efstathiou et al 2000,
Takagi et al 2003, Siebenmorgen \& Krugel 2007),
AGN dust tori (Rowan-Robinson \& Crawford 1989, Pier \& Krolik 1992, Granato \& Danese 1994, Efstathiou \& 
Rowan-Robinson 1995, Rowan-Robinson 1995, Nenkova et al 2002, 2008, Fritz et al 2006, H\"{o}nig et al 2006, 
Schartmann et al 2008), 
and quiescent ('cirrus') galaxies (Rowan-Robinson 1992, Silva et al 1998, Dale et al 2001, Efstathiou \& Rowan-Robinson 2003, 
Dullemond \& van Bemmel 2005, Piovan et al 2006, Draine  \& Li 2006, Efstathiou 
\& Siebenmorgen 2009).  Each of these model types involves at least two significant model parameters 
so there are a great wealth of possible models, particularly as a galaxy SED may be a mixture of all three types.

Rowan-Robinson and Efstathiou (2009) have shown how these models can be used to understand the
 interesting diagnostic diagram of Spoon et al (2007) for starburst and active galaxies, which
plots the strength of the 9.7$\mu$m silicate feature against the equivalent width of the 6.2$\mu$m PAH feature
for 180 starburst and active galaxies with {\it Spitzer} Infrared Spectrograph (IRS) spectra.  Increasing depth of the 9.7$\mu$m silicate feature is, 
broadly, a measure of the youth of the starburst,
because initially the starburst is deeply embedded in its parent molecular cloud.  The detailed starburst model of
Efstathiou et al (2000) shows the evolution of the starburst SED through the whole history of the starburst, from the
deeply embedded initial phase through to the Sedov expansion phase of the resulting supernovae.
However Rowan-Robinson and Efstathiou (2009) did find that there was some aliasing between young starbursts 
and heavily obscured AGN: the submillimetre data of {\it Herschel} can help to break this ambiguity, since young 
starbursts are expected to be much more prominent in the far infrared and submillimetre than AGN dust tori.

Often, however, we have only limited broad-band data available and in this situation it is more illuminating
to use a small number of infrared templates to match the observed infrared colours (eg Rowan-Robinson
and Crawford 1989, Rowan-Robinson 1992, 2001, Rowan-Robinson and Efstathiou 1993,
Rowan-Robinson et al 2004, 2005, 2008, Franceschini et al 2005, Polletta et al 2007, Magdis et al 2012).   
A set of just four templates (a quiescent 'cirrus' component, M82- and
Arp 220-like starbursts, and an AGN dust torus model) 
have proved remarkably successful in matching observed {\it ISO} and {\it Spitzer} SEDs, including cases where 
{\it Spitzer} IRS data are 
available (Rowan-Robinson et al 2006, Farrah et al 2008, Hernan-Caballero et al 2009).

\begin{figure}
\includegraphics[width=7cm]{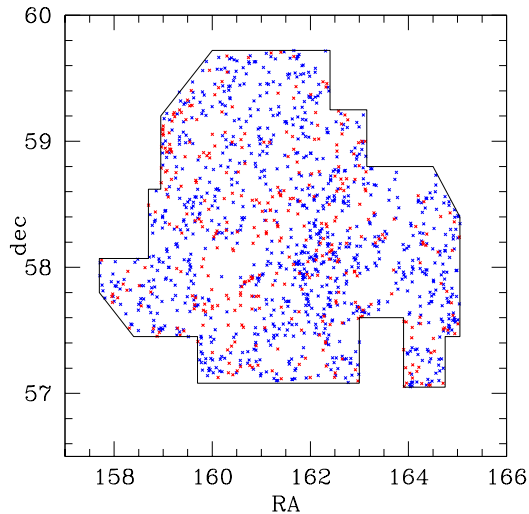}
\caption{
The region of overlap of the HerMES SPIRE survey area and the SWIRE photometric redshift survey in the Lockman Hole area.
HerMES 500$\mu$m sources associated with SWIRE sources in blue, unassociated sources in red.
}
\end{figure}

\begin{figure*}
\includegraphics[width=7cm]{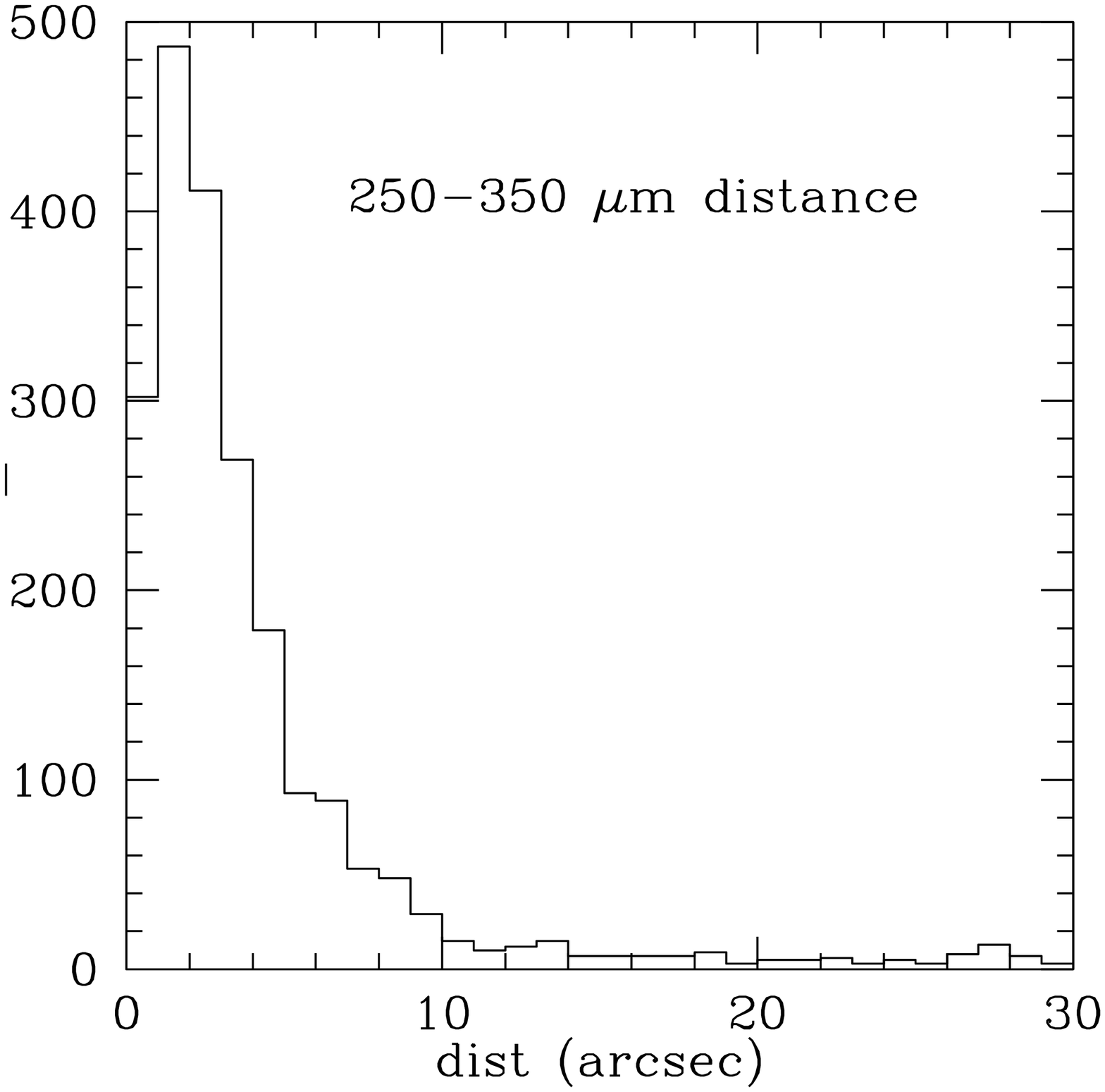}
\includegraphics[width=7cm]{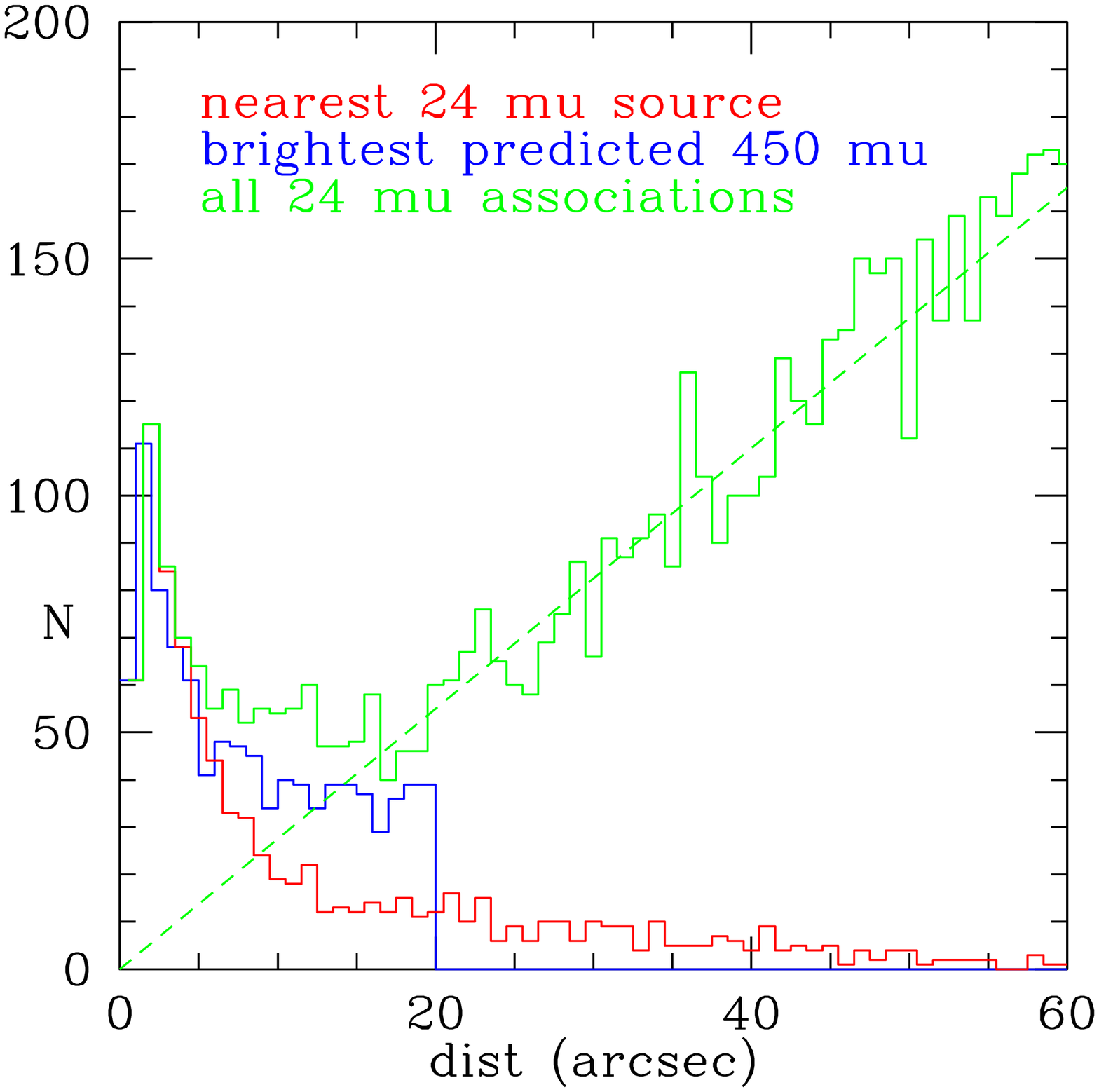}
\caption{LH: Distribution of separations between SCAT 250 and 350$\mu$m associations.
RH: Distribution of separations between SWIRE and SPIRE sources for nearest 24$\mu$m source (red),
all 24$\mu$m sources (green), and the brightest predicted 500$\mu$m source (blue).
}
\end{figure*}

\begin{figure}
\includegraphics[width=7cm]{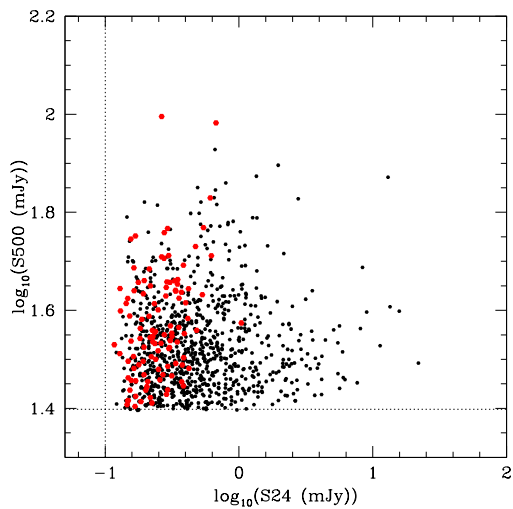}
\caption{
S500 versus S24 for Lockman-SCAT sources (red points: lens candidates, see section 4).  The dotted lines indicate the
500$\mu$m sensitivity limit of 25 mJy and the 24$\mu$m cutoff of 100$\mu$Jy.
}
\end{figure}

Following the discussion by Rowan-Robinson and Efstathiou (2009) of the Spoon et al (2007) Spitzer-IRS diagnostic diagram, 
and our modelling of a sample of 68 HerMES sources (Rowan-Robinson et al 2010), we
have introduced two new templates in our automated fits to far infrared and submillimetre data: a young (t=0) $A_V$ = 100 starburst (our M82 and A220 starburst templates correspond to older starbursts with
$A_V$ = 50, t =40 Myr, and $A_V$ = 200, t = 30 Myr, respectively)
and a cooler ($\psi$ = 1) cirrus model, where $\psi$ is the ratio of the intensity of the radiation field to that in the
solar neighbourhood (Rowan-Robinson 1992, Rowan-Robinson et al et al 2013: see section 3 below).  Our standard cirrus model corresponds to the ratio of the intensity of the radiation field to that in the solar neighbourhood, $\psi$ = 5.
In the present paper we explore whether this simple six-template approach works for galaxies detected
by the SPIRE array (Griffin et al 2010) on {\it Herschel}, and what additional infrared components may be present. 
We apply the method to a large sample of  {\it Herschel} sources, $\sim$ 1000 galaxies, in an area of the sky where we have
a wealth of ancillary data at optical, near-, mid- and far-infrared wavelengths.  We use physical models derived from radiative
transfer codes to gain a deeper understanding of the evolutionary status of the infrared galaxies and the balance between
starburst and quiescent phases.  Submillimetre wavelength data reveal the presence of colder dust than is detected at far
infrared wavelengths.  

Another key phenomenon in understanding submillimetre galaxies is gravitational lensing.  Negrello et al (2007, 2011) have shown 
that a significant fraction of bright 500 $\mu$m galaxies are lensed (see also Wardlow et al 2013).  In this paper we
explore whether anomalous excess submillimetre radiation in fainter galaxies may be due to lensing.
Detailed SED modelling is needed to distinguish cold dust from the effects of gravitational lensing.

A flat cosmological model with $\Lambda$=0.7, $h_0$=0.72 has been used throughout.

\section{Selection of sample with high quality flux-densities at 24, 250, 350 and 500$\mu$m}

In this analysis we have focused on the HerMES\footnote{HerMES.sussex.ac.uk}  (Oliver et al 2012) SCAT catalogue 
(Wang et al 2014) in the SWIRE-Lockman area (Lonsdale et al 2003), where we  have optical 
and 3.6-160$\mu$m {\it Spitzer} photometry, photometric redshifts, and infrared template fits from the SWIRE photometric 
redshift catalogue (Rowan-Robinson et al 2008, 2013).  

Our starting point is the HerMES SPIRE Catalogue (SCAT) 500$\mu$m catalogue (Wang et al 2014) with sources detected
at 500$\mu$m without using any prior information from the other SPIRE bands (i.e. "blind" 500$\mu$m catalogue).
We choose this starting-point, rather than the XID catalogue of Roseboom et al (2010), because we want to reconsider
the process of associating SPIRE sources with Spitzer 24 $\mu$m sources.
There are 2970 sources in the HerMES Lockman area that
are detected at better than 5-$\sigma$ at 500$\mu$m.  Here we are using the estimated total error, which includes
the contribution of confusion noise, $\sim$6.8 mJy$/$beam (Nguyen et al 2010).  The process of building a band-merged catalogue 
begins by looking for associations with 5-$\sigma$ 350$\mu$m 
SCAT sources, using a search radius of 30 arcsec.  The SPIRE beam full-width to half maximum is 18.2, 25.2, 36.3 arcsec at
250, 350, 500$\mu$m, respectively (Griffin et al 2013).  2709 sources found associations and we shall ignore the 261 
sources (8.7$\%$) which did not find associations
with well detected 350$\mu$m sources.  Some of these may be high redshift 'red' sources (Dowell et al 2014) while others may
be combinations of two or more fainter 500$\mu$m sources.  We note that 122 did find associations with 350$\mu$m sources detected at 4-5 $\sigma$
but we have not pursued these further here. Of these 2709 5-$\sigma$ 350-500$\mu$m sources, 
1335 lie in the area covered by the {\it Spitzer} SWIRE survey (Fig 1) and it is this latter sample that we focus on in this paper.

We next associated the 1335 350-500$\mu$m sources in the SWIRE area with 250$\mu$m sources detected at better than 5-$\sigma$,
using a search radius of 30 arcsec from the 350$\mu$m position.  879 found associations, and a further 163 found associations with 250$\mu$m
 sources detected at 4-5 $\sigma$.  The remaining 293 350-500$\mu$m sources we regard as undetected at 250$\mu$m.  Figure 2L
shows the distribution of separations between SCAT 250 and 350$\mu$m sources.  Most associations have separations $<$10
arcsec (97$\%$ are within 20 arcsec) and associations with separations in the range 20-30 arcsec should perhaps be treated with caution.

\begin{figure*}
\includegraphics[width=7cm]{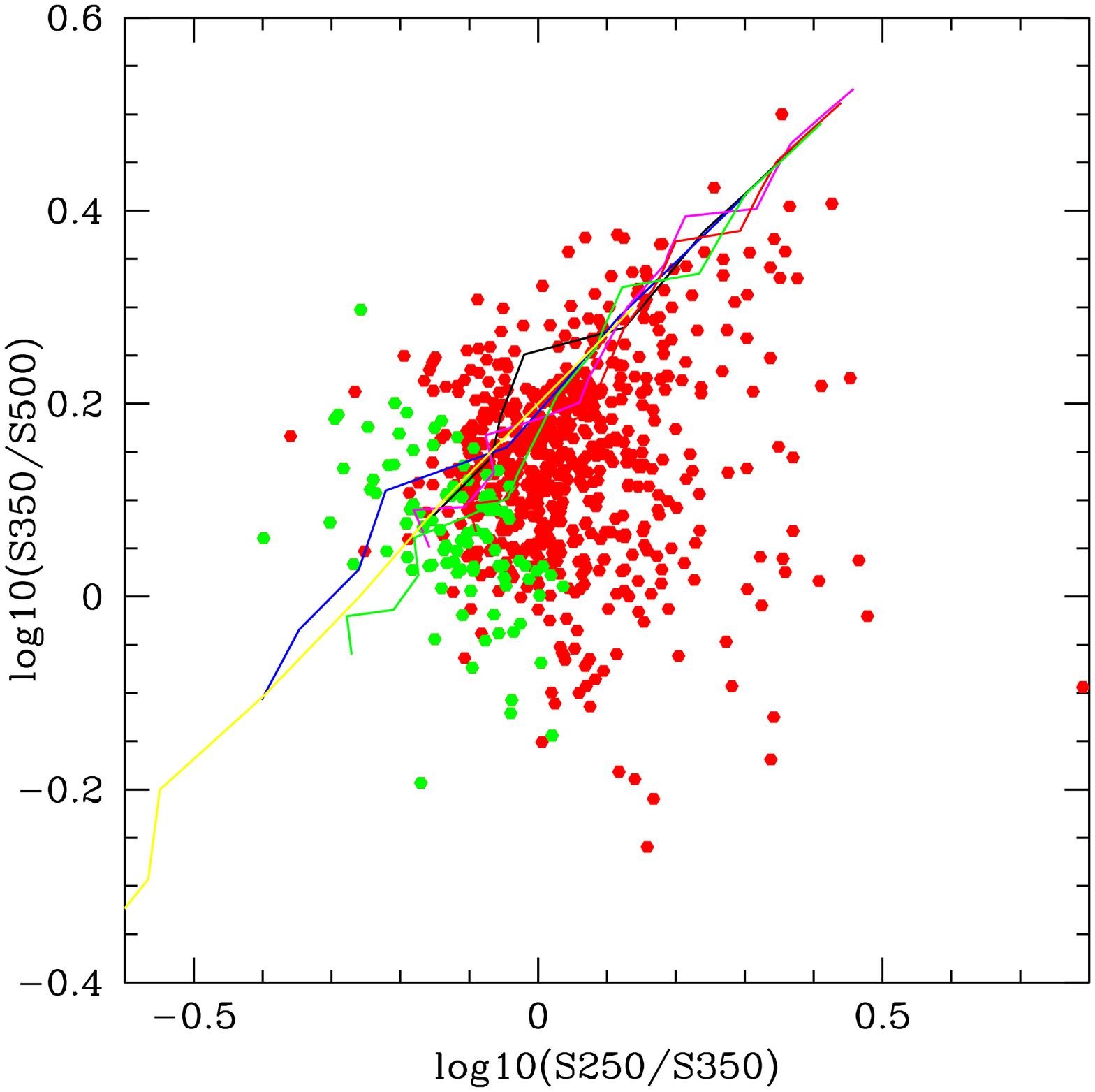}
\includegraphics[width=7cm]{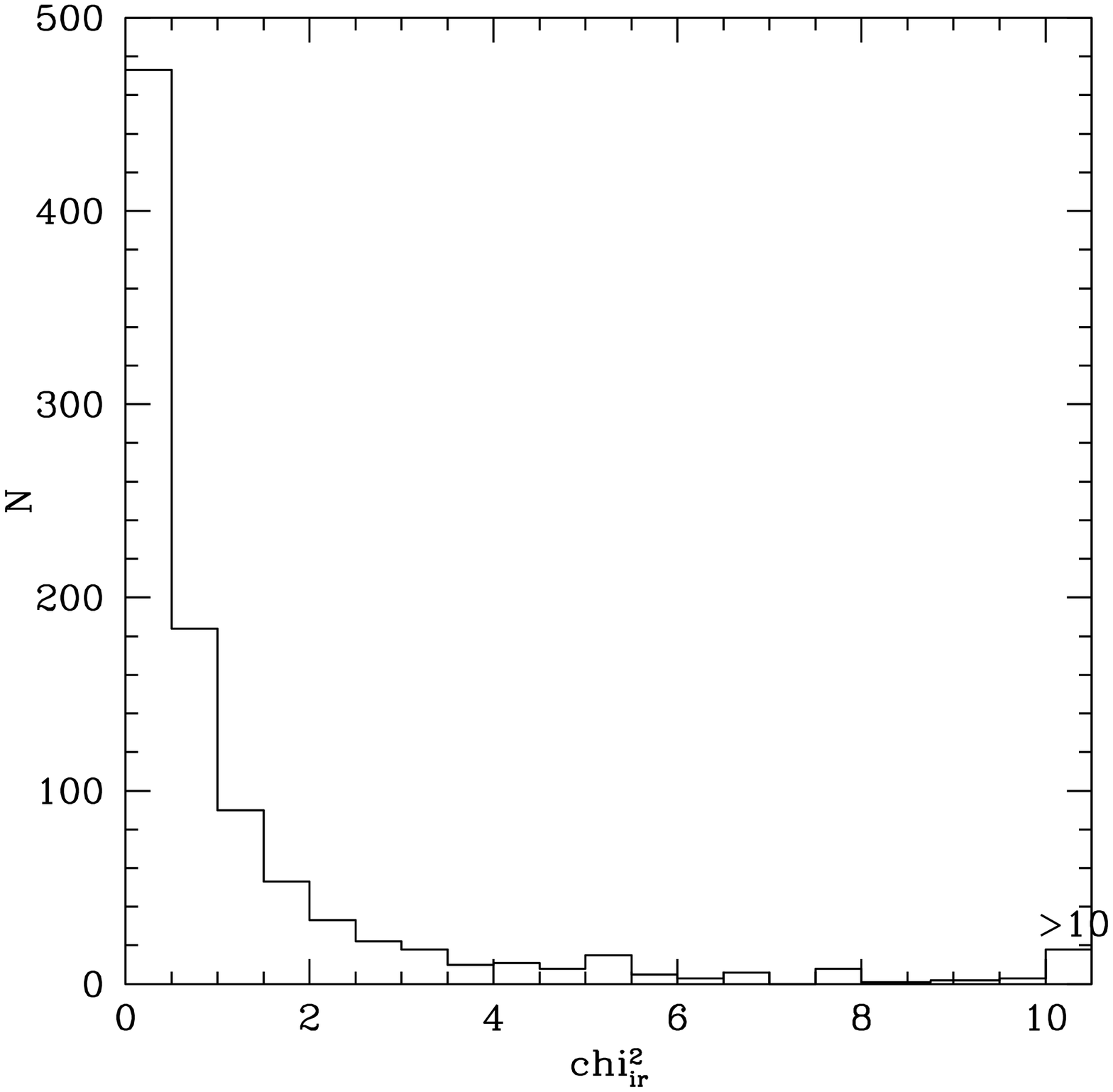}
\caption{LH:250-350-500$\mu$m colour-colour diagram for SCAT Lockman sources with 5-$\sigma$ detections at 250
and 350$\mu$m, and which are also associated with SWIRE 24$\mu$m sources (red: $> 5-\sigma$ at 250$\mu$m;
green: $4-5-\sigma$ at 250$\mu$m.  Coloured loci are for different templates, with z = 0 at top right, z = 4 at bottom left.
RH: Histogram of reduced $\chi^2_{\rm ir}$ from our automated 6-template fit to our sample of 967 HerMES-Lockman galaxies.
}
\end{figure*}

Finally we associated the 1335 350-500$\mu$m sources with the SWIRE Photometric Redshift Catalogue (Rowan-Robinson et al 2008, 2013),
using a 20 arcsec search radius from the 250$\mu$m position (or 350$\mu$m if no 250$\mu$m detection), and requiring a 24$\mu$m
detection brighter than 100$\mu$Jy, above which the SWIRE 24$\mu$m catalogue is $\sim 95\%$ reliable.  The 24$\mu$m  70$\%$ completeness 
limit is $\sim$300$\mu$Jy in the Lockman field (90$\%$ at 400$\mu$Jy).

The SWIRE Photometric Redshift Catalogue contains over 1 million galaxies covering 49 sq deg of sky, including 7.53 sq deg in Lockman,
which had been surveyed at 3.6-160$\mu$m by {\it Spitzer} and at ugrizJHK in ground-based surveys.  Photometric redshifts were
determined from ugrizJHK and 3.6, 4.5$\mu$m photometry (with up to 13 photometric bands available in the Lockman area) 
and the 5.8-160$\mu$m data were fitted with infrared templates.  These
allowed prediction of fluxes at submilllimetre wavelengths.   If more than one SWIRE source is associated with a HerMES SCAT
source, the source with the highest predicted 450$\mu$m flux from the SWIRE Photometric Redshift Catalogue
is selected.  Thus the redshift and the 5.8-160$\mu$m flux information are being used to select the best association.  Inevitably
there will be cases where the submillimetre flux should be assigned to more than one source because of the effects of confusion,
and here a more sophisticated treatment would require comparison of multiwavelength maps.
However we believe our approach gives more reliable results than assigning flux purely on the basis of the 24$\mu$m flux, as in Roseboom
et al (2010).  We found 
a total of 967 SWIRE associations, 73$\%$ of the 1335 total.  For 13 sources two SPIRE sources selected the same 24$\mu$m
source and in these cases we deleted the weaker 500$\mu$m source from the list.  We expect the bulk of the 368 unassociated 
sources to be at redshift $>$1.5, given the SWIRE selection function.  Figure 3 shows 500 $\mu$m flux-density versus 24 $\mu$m flux-density 
for the associated sources.  The 24$\mu$m limit of 100 mJy has little effect on the sample selection.
Figure 4L shows the 250-350-500$\mu$m colour diagram with the loci of the infrared templates overplotted.  The lower SNR
250$\mu$m sources tend to have bluer SPIRE colours and be at higher redshift.

This is essentially a 500$\mu$m selected sample, with a flux-limit $\sim$25mJy, with the additional requirements of multi-wavelength associations.  The reason for 
making the selection at 500$\mu$m is to maximise the number of sources with photometry at
all 3 SPIRE wavelengths, since the focus of this paper is on SED modelling.
Our requirement of association with a SWIRE 24$\mu$m source discriminates against sources with
S(500)/S(24) $>$ 200, and our requirement of an entry in the SWIRE Photometric Redshift Catalogue discriminates
against sources with {\it z}$>$ 1.5 (see Fig 14R of Rowan-Robinson et al 2008).  Selection at 500$\mu$m favours 
galaxies with cooler dust, than say selection at 70 or even 250$\mu$m. 

Figure 2R shows the distribution of separations between SPIRE and SWIRE sources for the nearest 24$\mu$m source (in red),
for all 24$\mu$m sources, and for our adopted associations with the SWIRE 24$\mu$m source within 20 arcsec with the brightest 
predicted 500$\mu$m flux.  The green distribution shows the linear increase at large separations characteristic of random associations.
Our procedure ensures that the incidence of chance associations is kept low, but will be higher where the separation is $>$10 arcsec.  Although our 500 $\mu$m flux-density selection limit is set at five times the total error, including confusion noise, 
it is possible that a few 500 $\mu$m sources are in fact blends of two or more fainter sources
(see further discussion of the issues of confusion and misassociation in Section 5).

\section{Spectral energy distributions of \textit {Herschel} galaxies}
We first ran an automated infrared SED-fitting code to the 3.6-500$\mu$m (10 photometric band) data for all 967 HerMES-SWIRE 
sources in Lockman, 
using the six infrared templates discussed in section 1.  Figure 4R shows a histogram of the reduced $\chi^2_{\rm ir}$ for these fits.  The
fits are satisfactory except that there is a long tail of high $\chi^2_{\rm ir}$  values.
Fits where $\chi^2_{\rm ir} > 5$, of which there are 69 out of the 967 total, we regard as requiring further 
explanation, either in terms of photometry problems,
a need for additional infrared templates, or as indicating the presence of gravitational lensing.

However even where $\chi^2_{\rm ir} < 5$, we regard a solution where the luminosity in the optically thin 'cirrus' component is greater
than the total luminosity in starlight (after correction for extinction), as unphysical and requiring further scrutiny.

We have plotted here (Figs 5-14, 18-23) the individual SEDs of a large sample of the 967 HerMES-SWIRE sources in Lockman, 
259 sources in total.  These are divided
into four distinct subsamples: (1) sources with  $\chi^2_{\rm ir} > 5$ and at least 10 optical-nir (0.36-4.5$\mu$m) photometric bands, 
(2) sources with $\chi^2_{\rm ir} < 5$ and at least 12 optical-nir photometric bands, (3) sources with 10 photometric bands and $\chi^2_{\rm ir} < 5$ 
but $L_{\rm cirr} > L_{\rm opt}$, (4) candidate gravitational lenses (see section 4).

The individual SED modelling follows the methodology of Rowan-Robinson et al (2005, 2008), but with some new features.   
Optical and near infrared data are fitted with 
one of six galaxy templates and two QSO templates, with the extinction $A_V$ as a free parameter.  Infrared and submillimetre data are
fitted with a combination of seven infrared templates (three cirrus models ($\psi$ = 5, 1, 0.1, see below) , one of three starburst models (M82, A220
and a young starburst), and an AGN dust torus model).  The 'cool' cirrus template ($\psi$ = 1) and 'cold' cirrus template ($\psi$ = 0.1), and
the young starburst template are brought in following the demonstration by Rowan-Robinson et al (2010) of the need for these
templates when modelling {\it Herschel} sources.  Most SEDs are well-fitted by only one or two infrared templates, and in no case is more
than four templates used. 

Spectroscopic redshifts have been indicated in the SED plots by showing 4 decimal places.  The accuracy of the
photometric redshifts is $\sim 4\%$ in (1+z) for most of the galaxies, where 6 or more photometric bands are available
(Rowan-Robinson et al 2013).  The full $\chi^2$ distribution for each photometric redshift estimate is given in
the SWIRE Photometric Redshift Catalogue.
The optical galaxy templates are those of Rowan-Robinson et al (2008) and are shown at full resolution
in the SED plots.  

\begin{figure*}
\includegraphics[width=7cm]{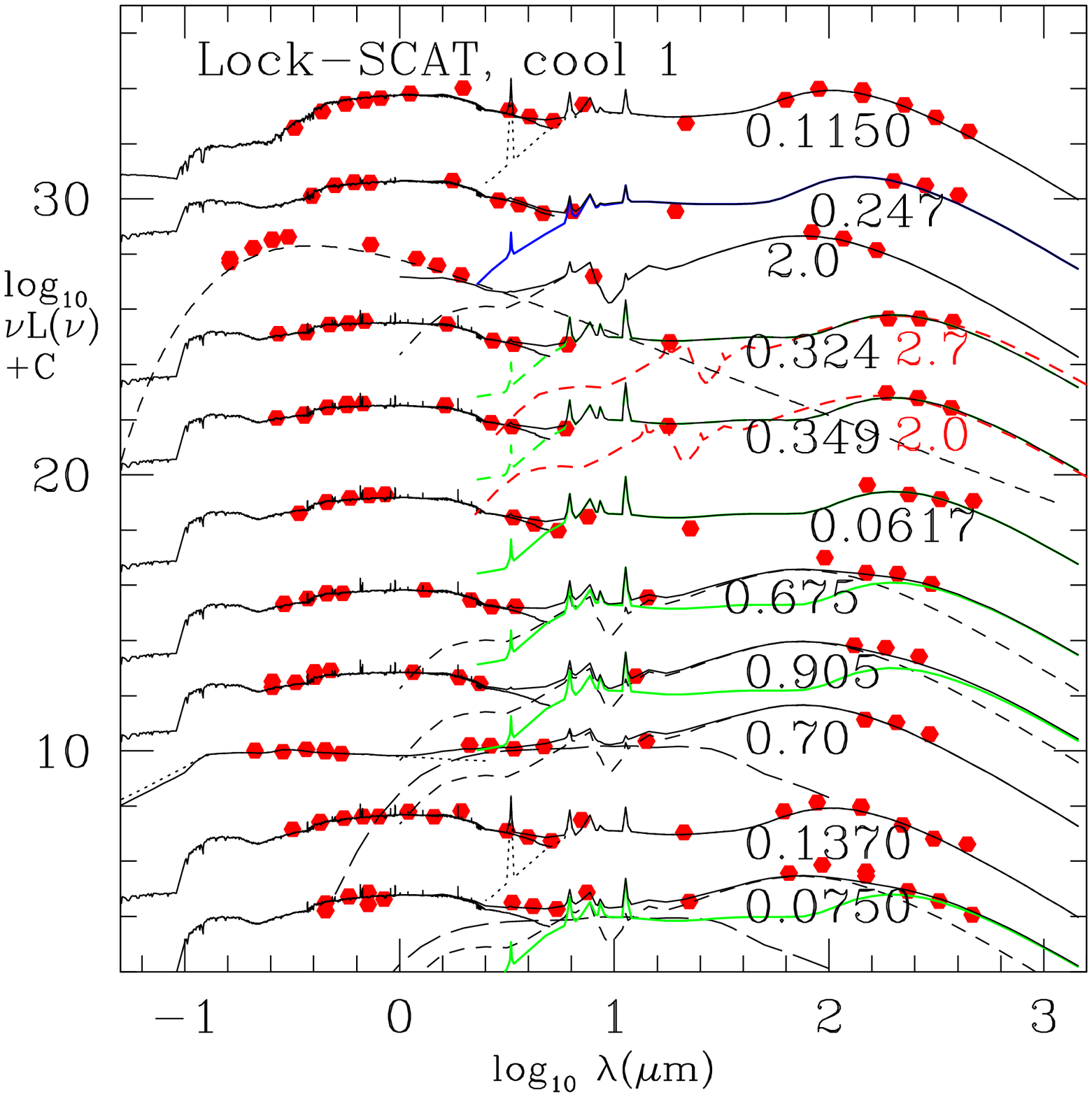}
\includegraphics[width=7cm]{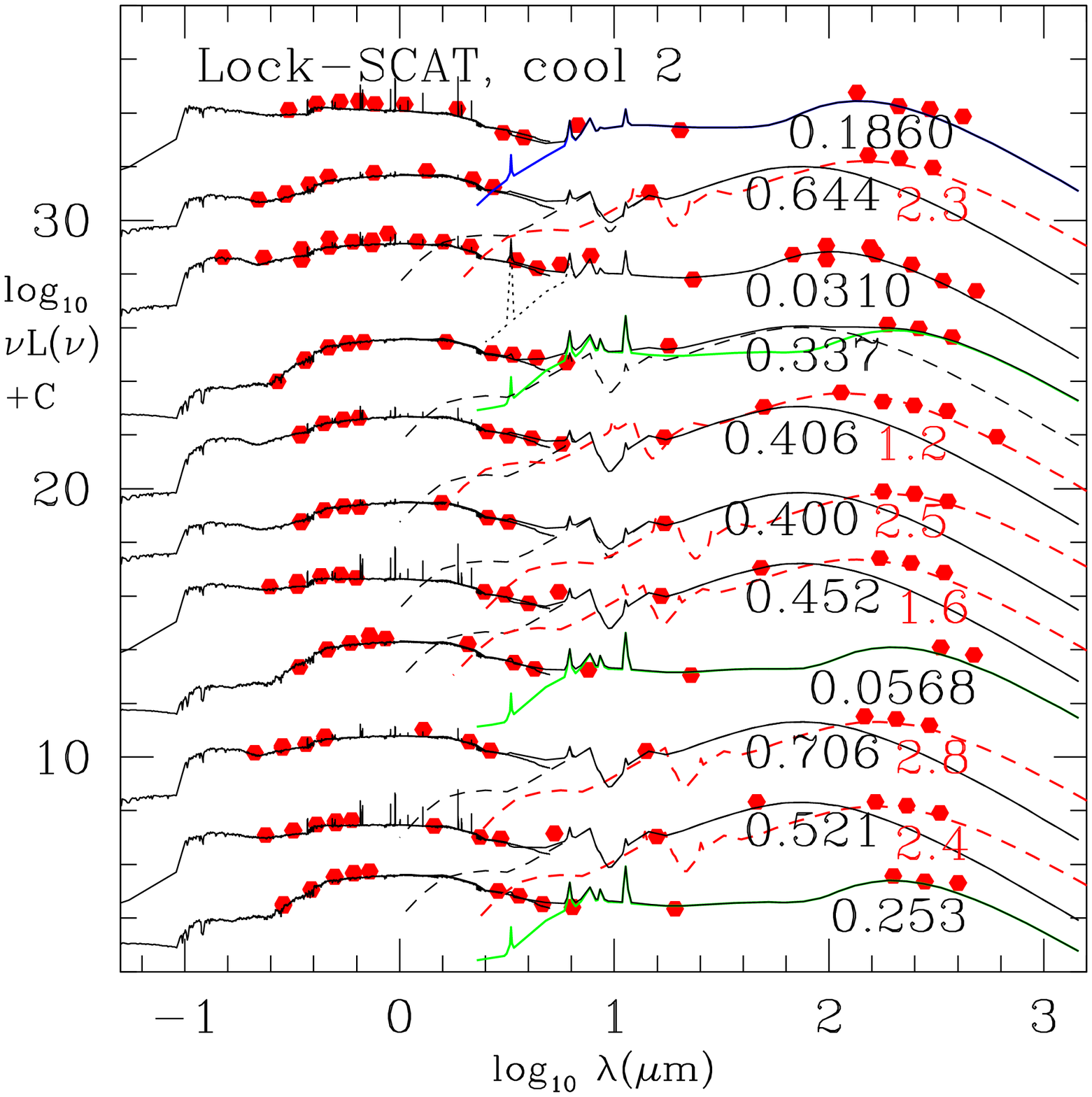}
\caption{SEDs for SWIRE-Lockman galaxies with at least 10 optical-nir photometric bands and with $\chi^2_{\rm ir} >$ 5.
Cool cirrus models ($\psi$=1) are shown in blue, cold cirrus models ($\psi$=0.1) in green.
Lens models are shown as red broken curves, with associated redshift of lensed galaxy in red.
Black dotted curves are cirrus ($\psi$=5) or M82 starbursts, black dashed curves are Arp 220 starbursts
and black long-dashed curves are AGN dust tori.
Where a lensed galaxy is also shown with a cirrus template fit, the latter has been rejected as
physically implausible because $L_{\rm cirr} > L_{\rm opt}$.
}
\end{figure*}

\begin{figure*}
\includegraphics[width=7cm]{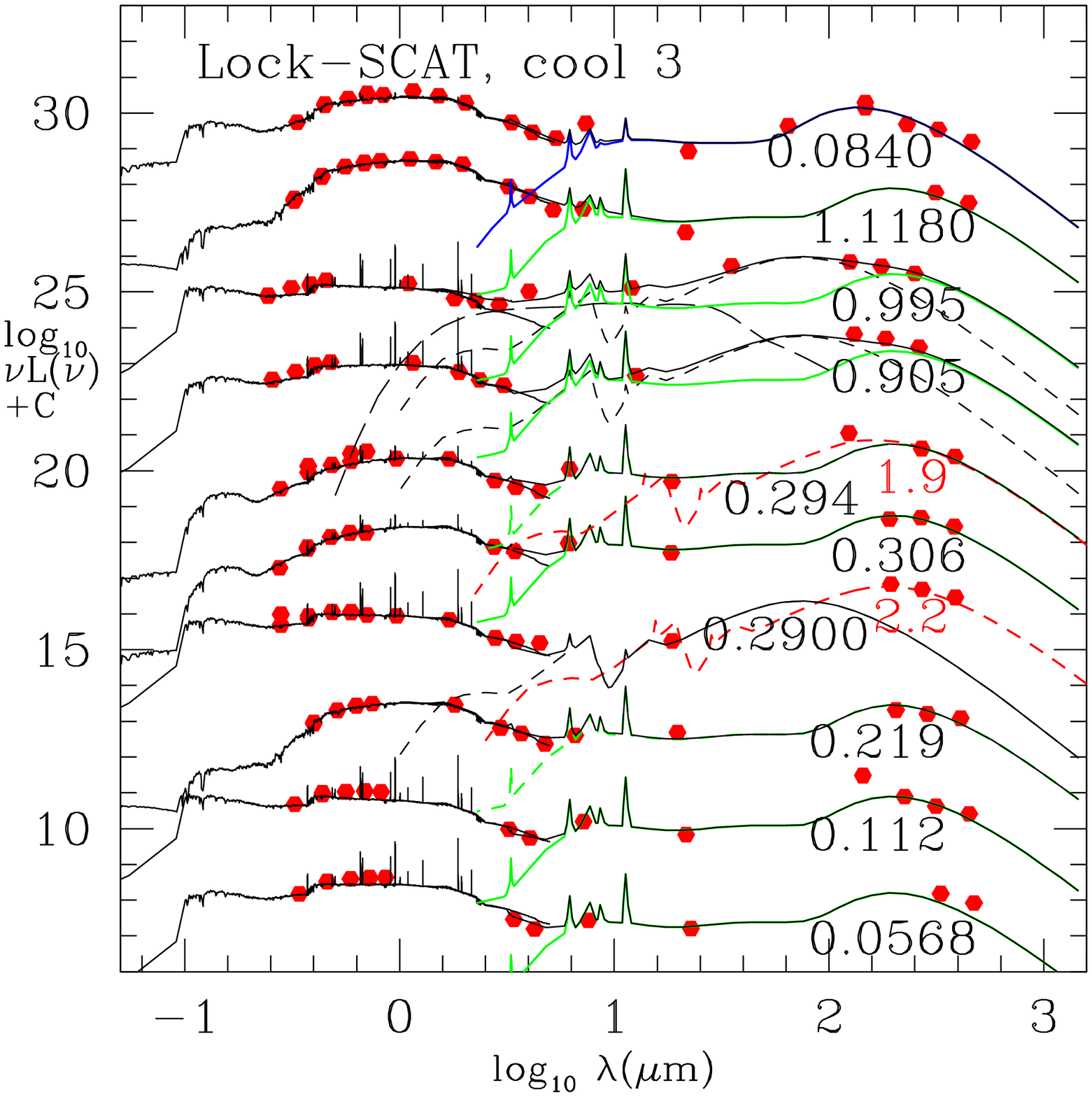}
\includegraphics[width=7cm]{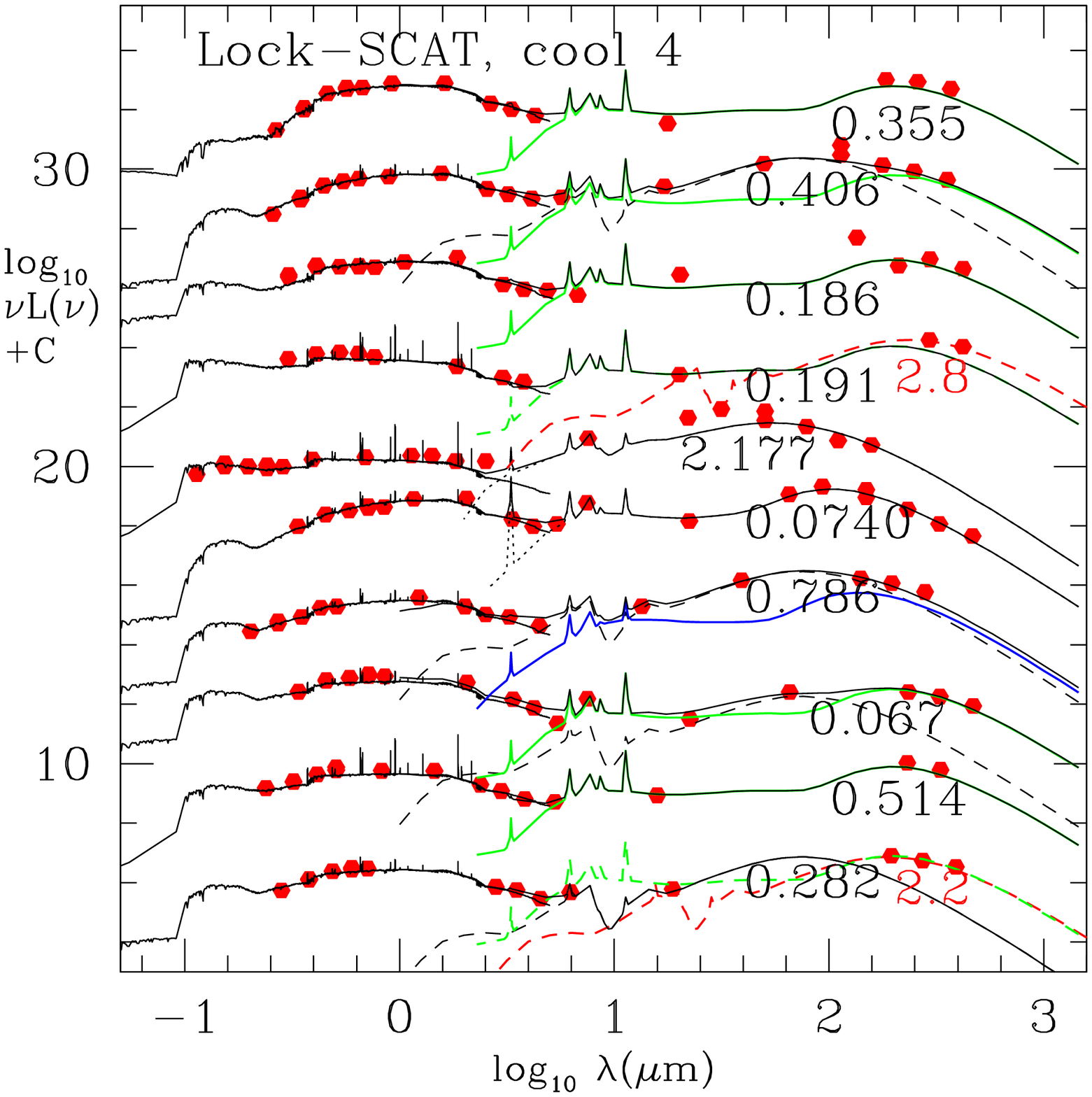}
\caption{SEDs for SWIRE-Lockman 10-band galaxies with $\chi^2_{\rm ir} >$ 5. Colour coding as in Fig 5. 
}
\end{figure*}

\begin{figure*}
\includegraphics[width=7cm]{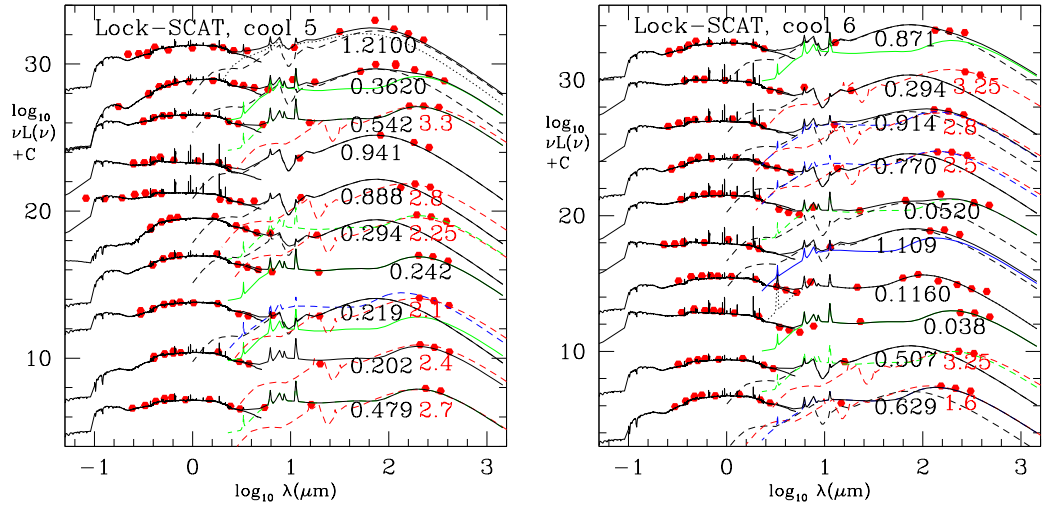}
\includegraphics[width=7cm]{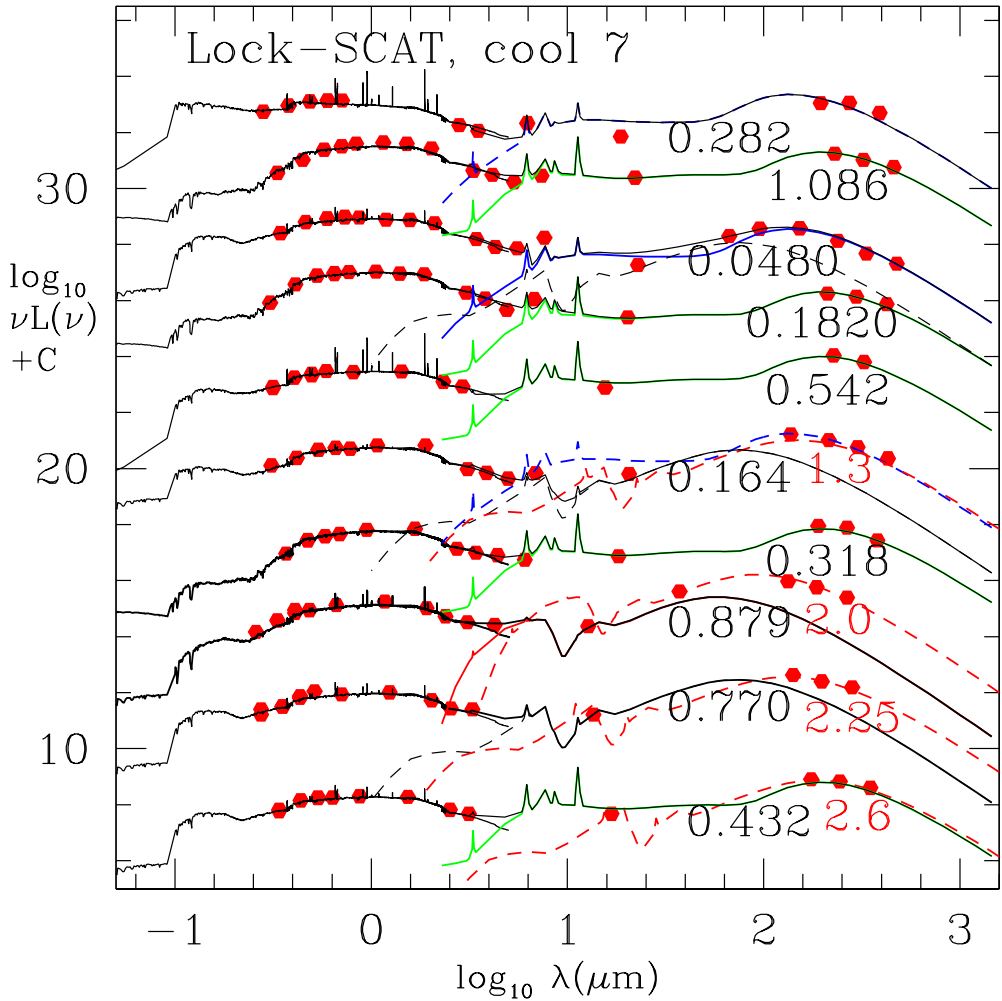}
\caption{SEDs for SWIRE-Lockman 10-band galaxies with $\chi^2_{\rm ir} >$ 5.. 
}
\end{figure*}

\begin{figure*}
\includegraphics[width=14cm]{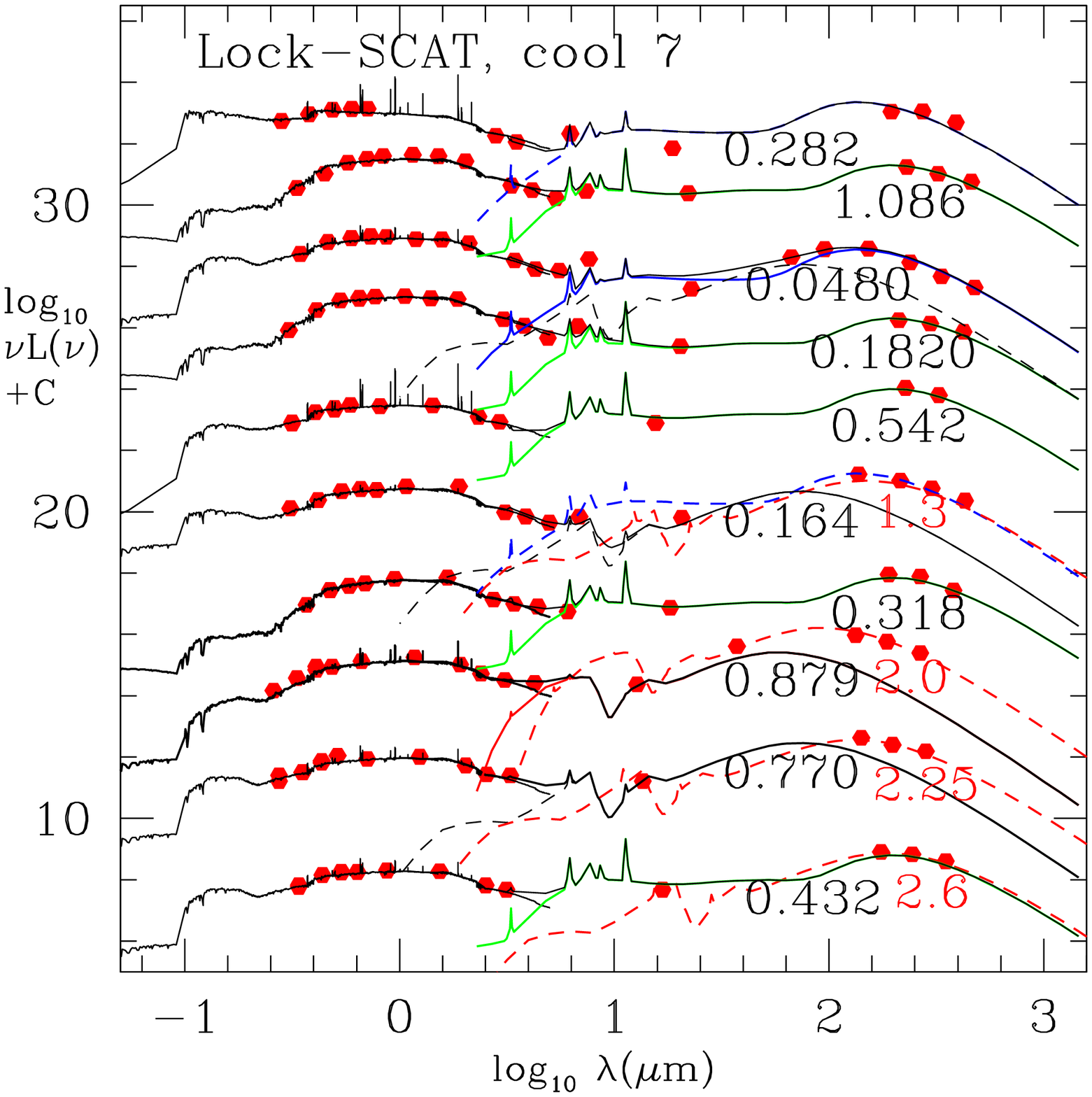}
\caption{SEDs for SWIRE-Lockman 10-band galaxies with $\chi^2_{\rm ir} >$ 5.. 
}
\end{figure*}

While the four standard infrared templates work well for many sources, the 350 and 500 
$\mu$m fluxes often require the presence of colder dust than is incorporated into our four basic templates.  
The two new cirrus templates used here are taken from the range of optically thin interstellar medium ('cirrus') templates developed by 
Rowan-Robinson (1992) and Efstathiou and Rowan-Robinson (2003).  The key parameter determining the temperature of the 
dust grains for optically thin emission is the intensity of the radiation field, which we can characterize by the ratio of intensity of 
radiation field to the local Solar Neighbourhood interstellar radiation field, $\psi$.  The standard cirrus template corresponds 
to $\psi$ = 5, and this
is the value used by Rowan-Robinson (1992) to fit the central regions of our Galaxy.  $\psi$ = 1 corresponds to the interstellar
radiation field in the vicinity of the Sun.  We also find that some galaxies need a much lower intensity radiation field than this,
with $\psi$ = 0.1.  The corresponding grain temperatures in the dust model of Rowan-Robinson (1992) are given in Table 1
of Rowan-Robinson et al (2010).  For the
two new templates, the ranges of dust grain temperatures for the different grain types are 14.5-19.7 K and 9.8-13.4 K respectively.  
Full details of the templates used are given in a readme page\footnote
{http:$//$astro.ic.ac.uk$/$public$/$mrr$/$swirephotzcat$/$templates$/$readme}.

The need for cooler dust templates can also be seen clearly in a plot of S(500)/S(24) versus
redshift (Fig 16), in which the predictions of different templates are shown.  At  {\it z} $<$ 1, a significant 
fraction of galaxies require colder dust than the standard cirrus model.  Hints of this population were seen at {\it z} $<$ 0.4 in
the plot of {\it ISO} 175$/$90$\mu$m flux ratio versus redshift (Fig 23) of Rowan-Robinson et al (2004).  Symeonidis et al (2009)
plotted a very similar figure, 160$/$70$\mu$m flux ratio versus redshift, for {\it Spitzer} data.  They interpreted this as implying 
strong evolution in the cold dust component.  The need for cooler dust was also seen in the Planck study of nearby galaxies
(Ade et al 2011).

\begin{figure*}
\includegraphics[width=7cm]{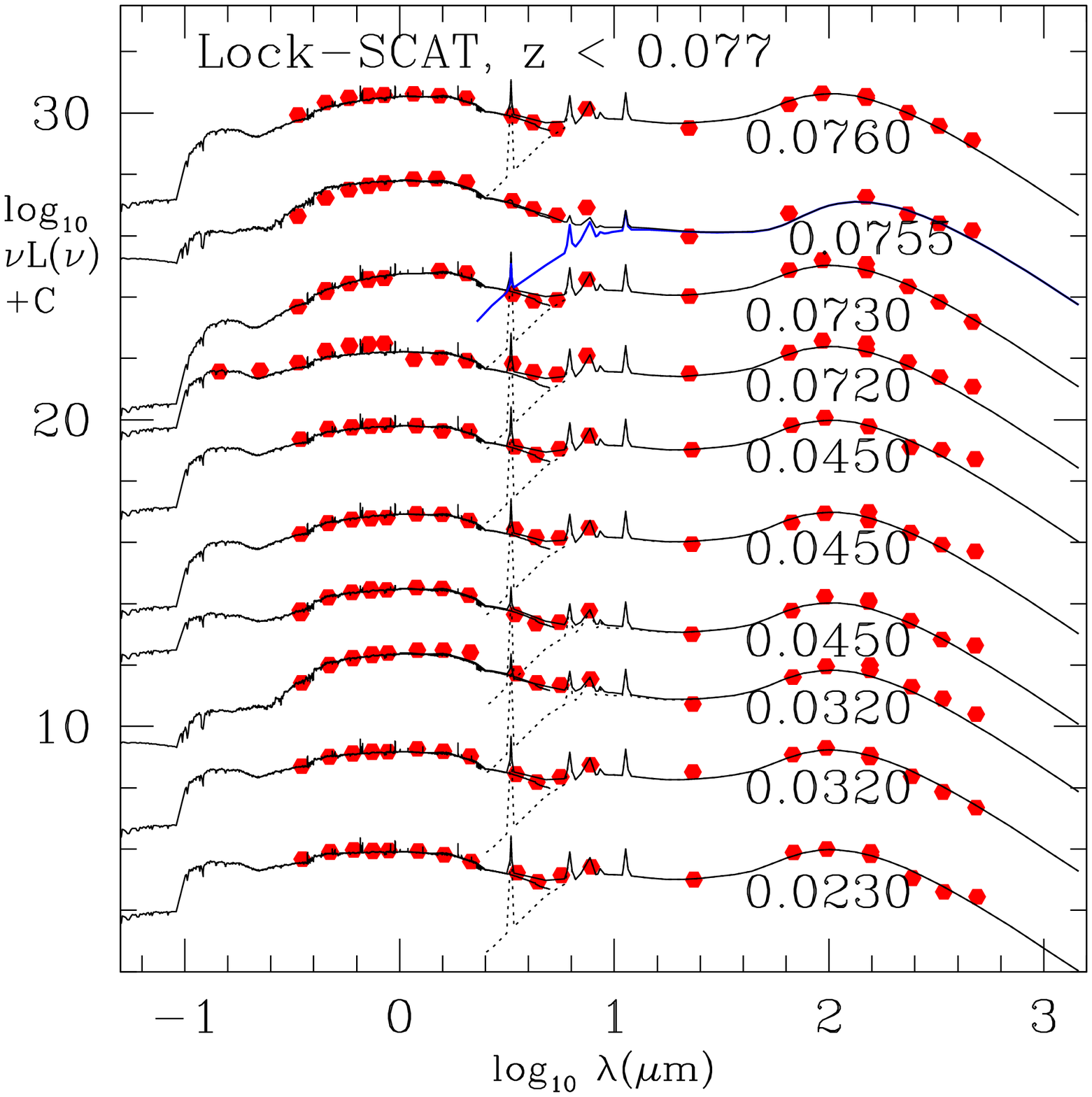}
\includegraphics[width=7cm]{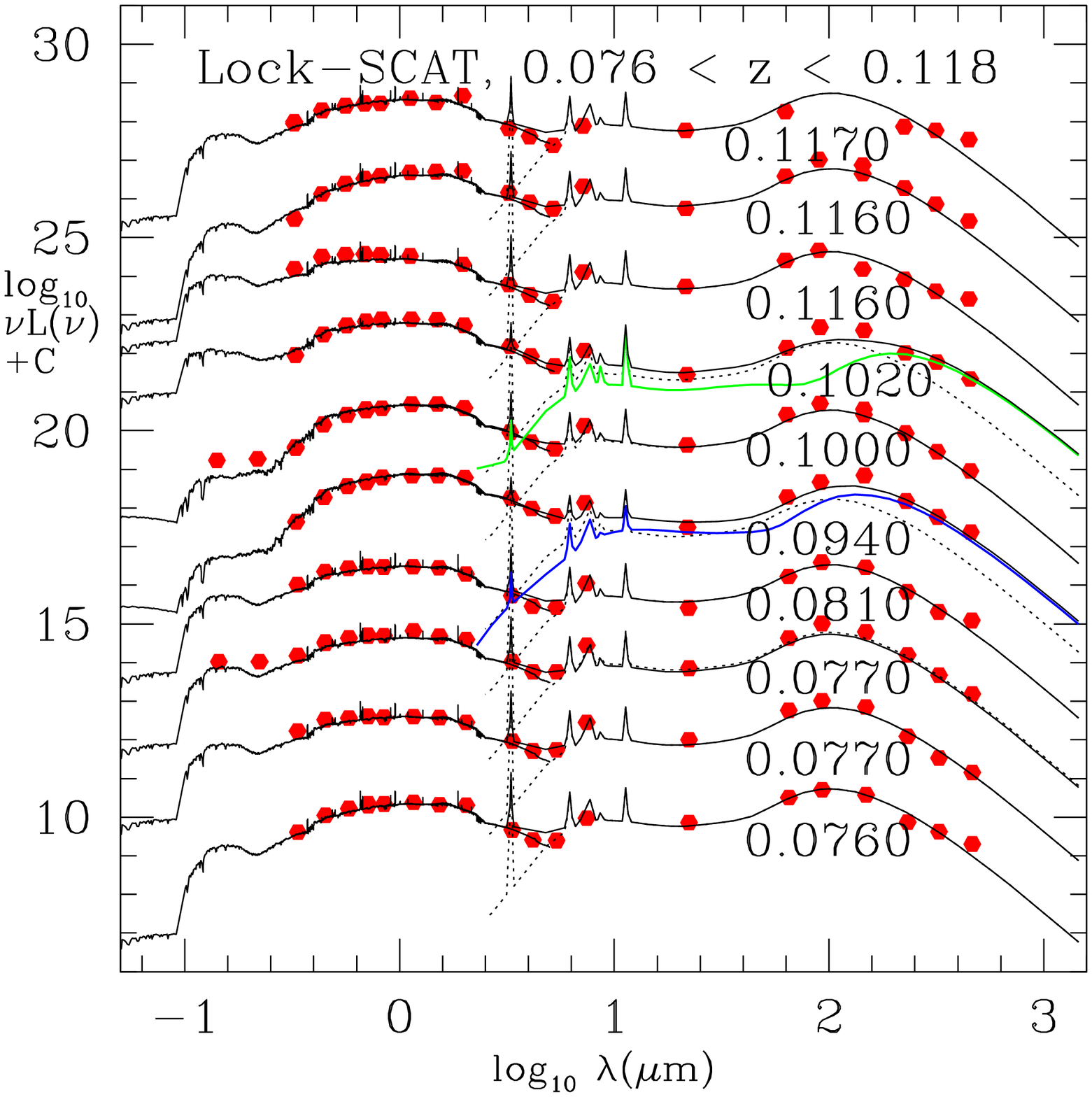}
\caption{L: SEDs for SWIRE-Lockman galaxies with at least 12 optical-nir photometric bands and with {\it z}$<$ 0.077.  
Photometric redshifts are indicated with only 3 significant figures.   Colour-coding as in Fig. 5.
R: SEDs for SWIRE-Lockman 12-band galaxies with 0.076 $<$ {\it z}$<$ 0.118. 
}
\end{figure*}

\begin{figure*}
\includegraphics[width=7cm]{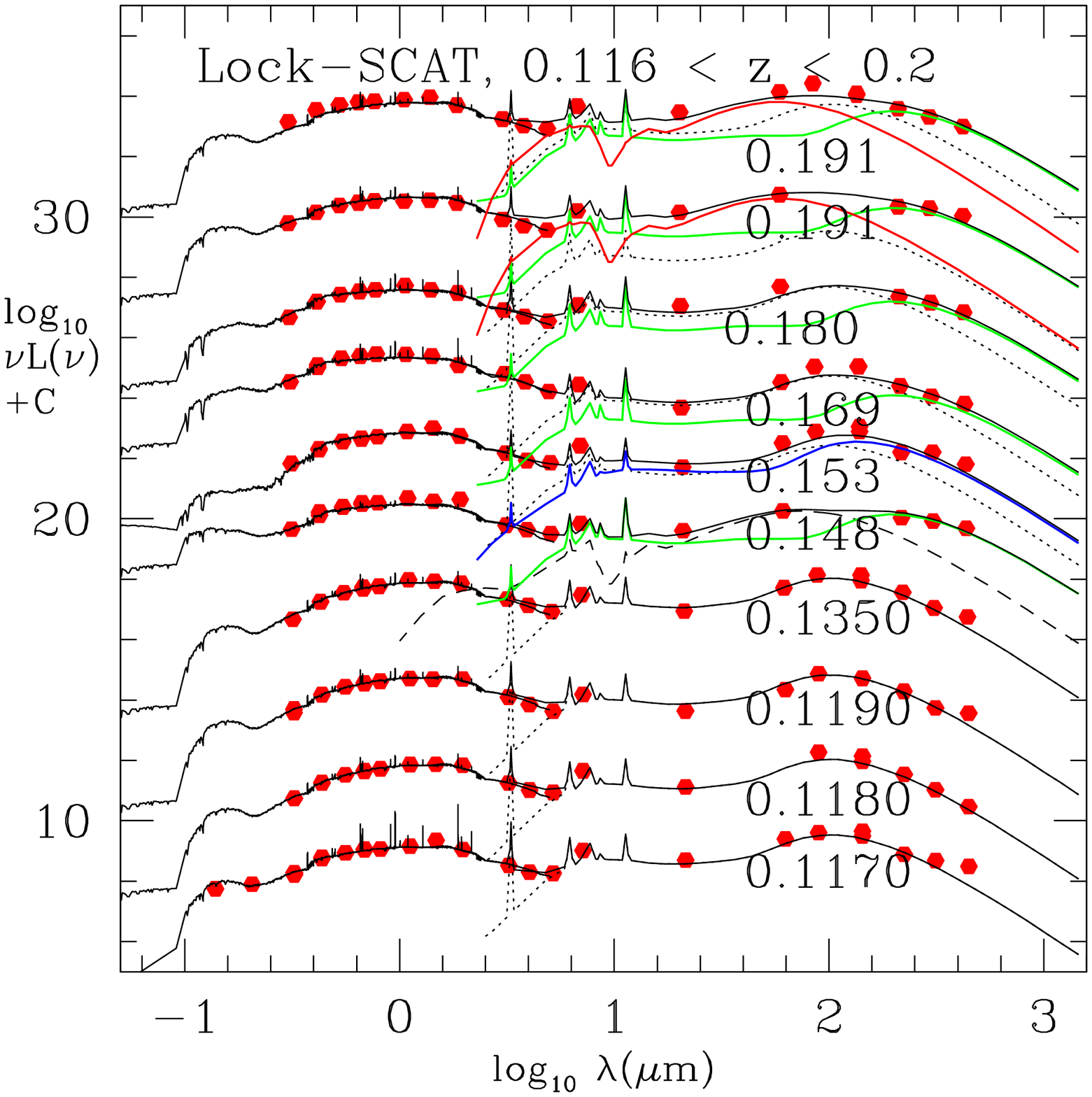}
\includegraphics[width=7cm]{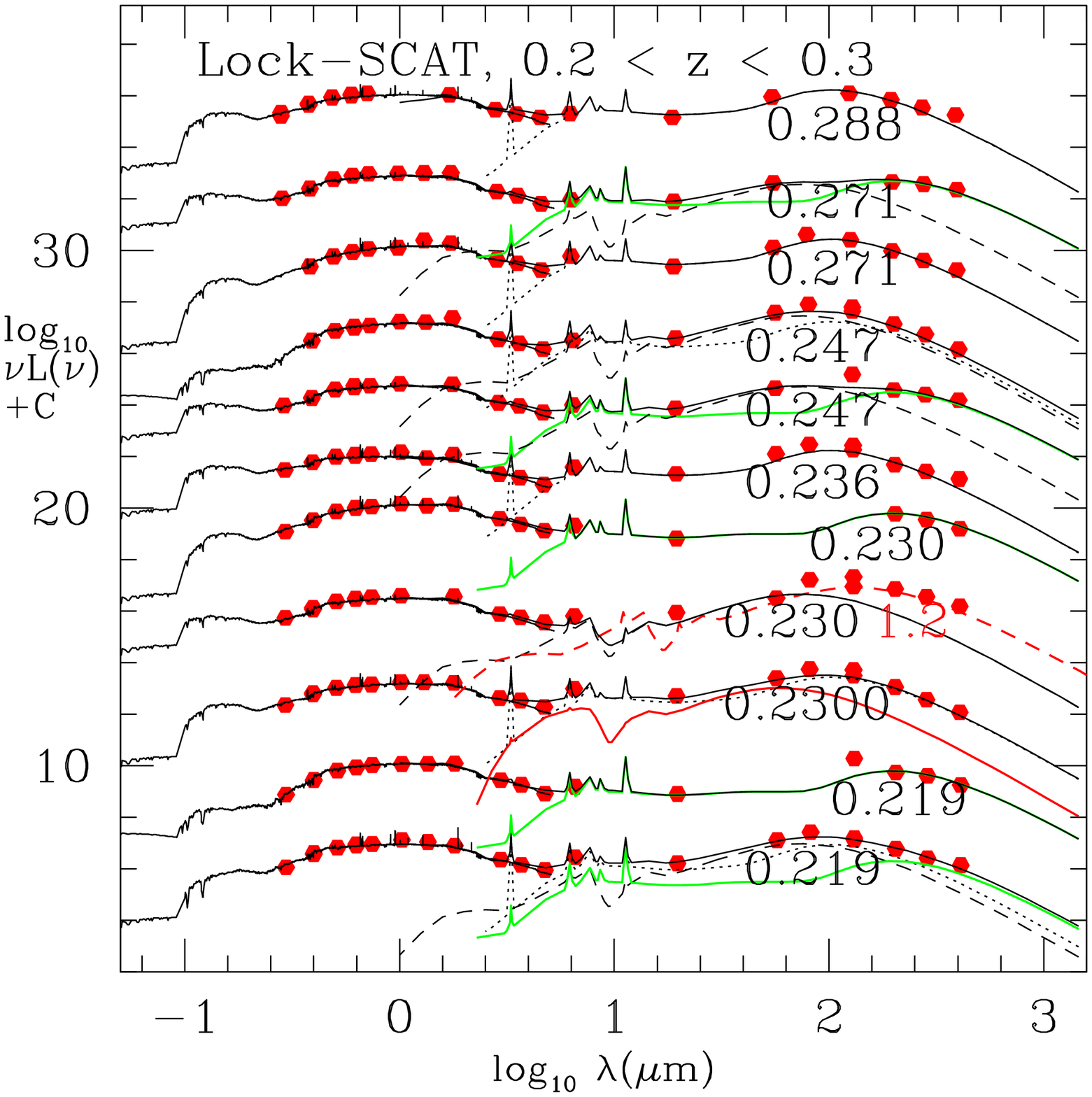}
\caption{L: SEDs for SWIRE-Lockman 12-band galaxies with 0.116 $<$ {\it z}$<$ 0.2. 
R: SEDs for SWIRE-Lockman 12-band galaxies with 0.2 $<$ {\it z}$<$ 0.3.
Solid red curves are young starbursts.
}
\end{figure*}

\begin{figure*}
\includegraphics[width=7cm]{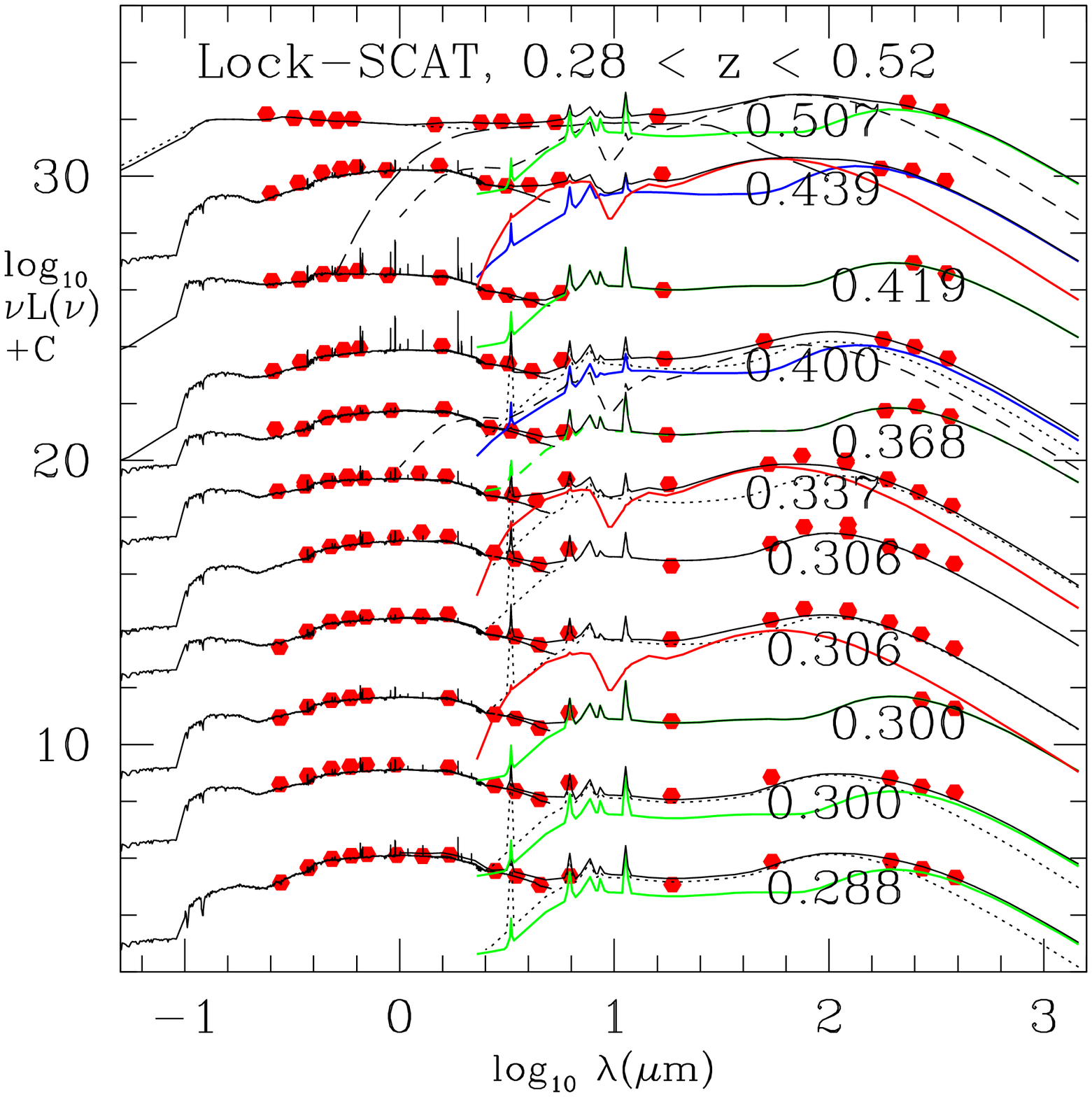}
\includegraphics[width=7cm]{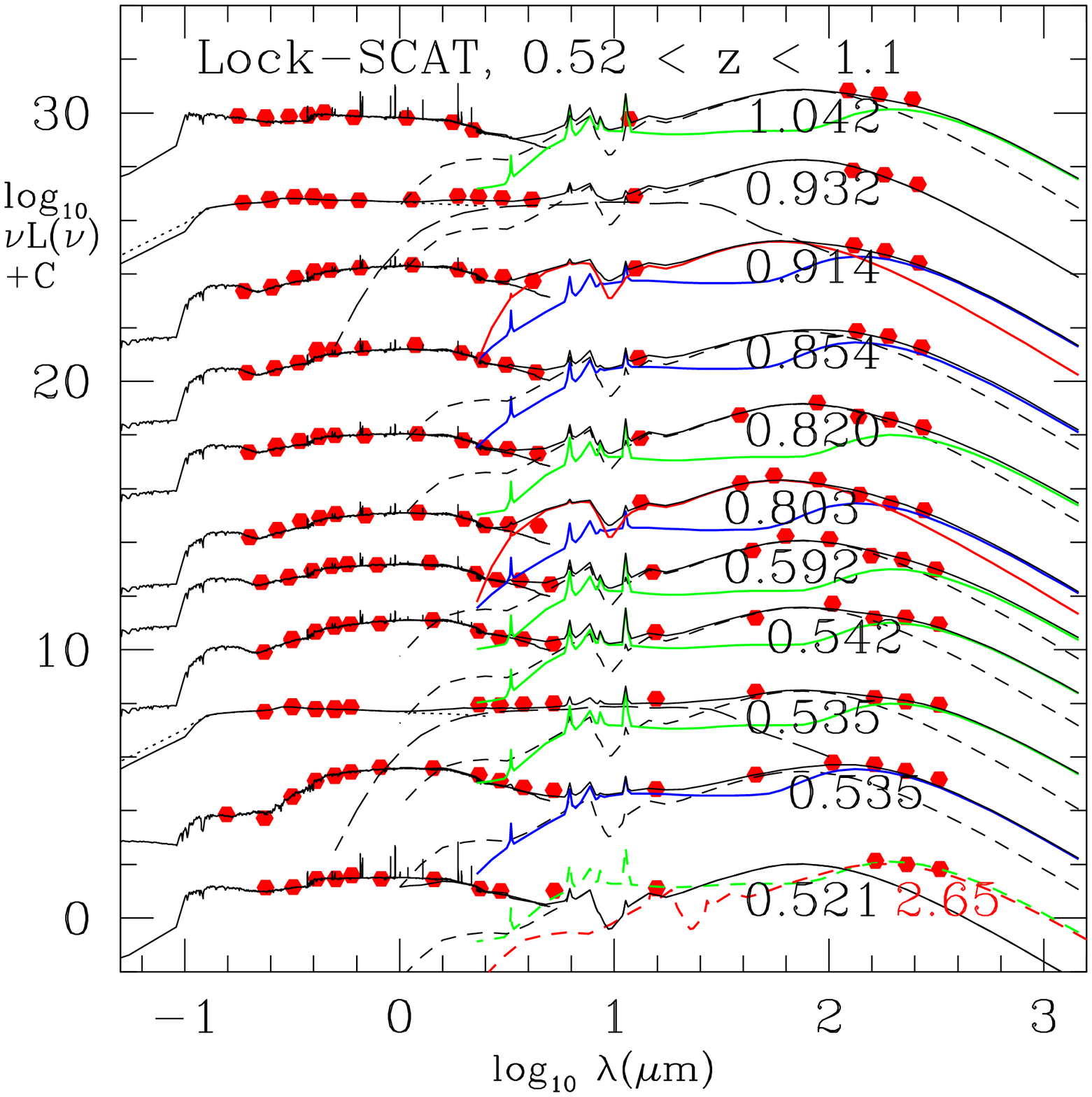}
\caption{L: SEDs for SWIRE-Lockman 12-band galaxies with 0.28 $<$ {\it z}$<$ 0.52. 
R: SEDs for SWIRE-Lockman 12-band galaxies with 0.52 $<$ {\it z}$<$ 1.1. 
}
\end{figure*}

\begin{figure*}
\includegraphics[width=7cm]{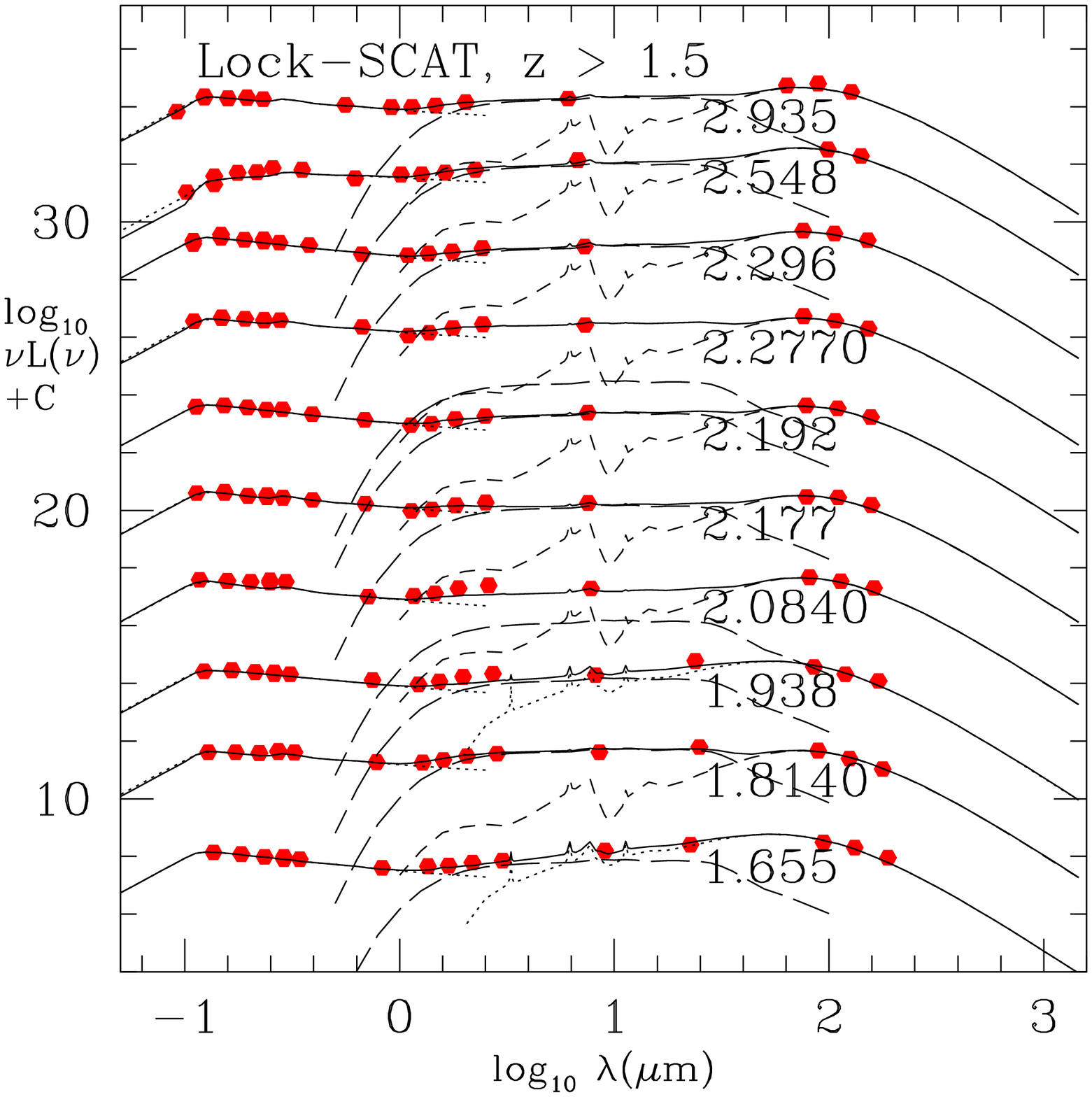}
\includegraphics[width=7cm]{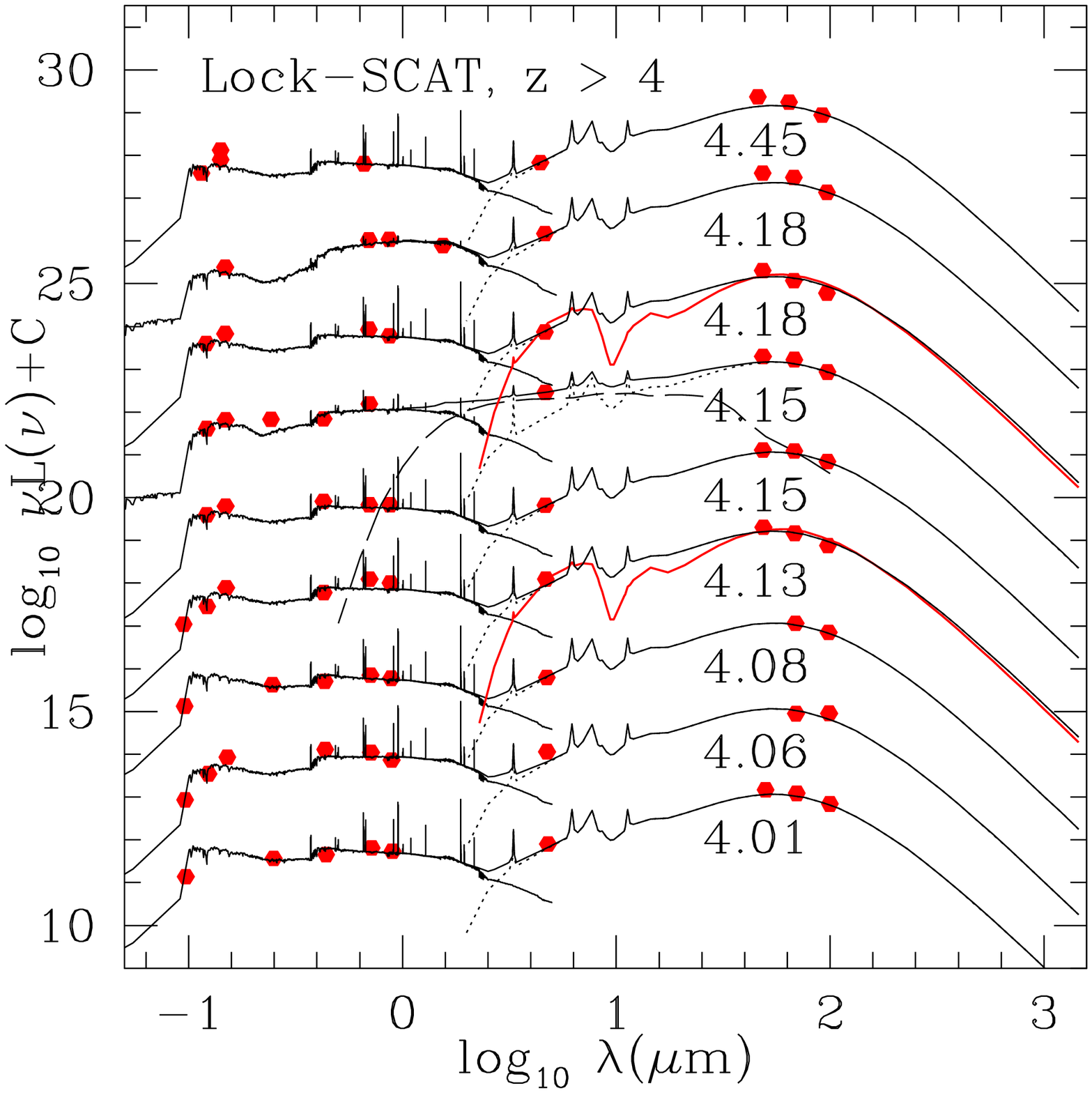}
\caption{L: SEDs for SWIRE-Lockman 12-band galaxies with {\it z}$>$ 1.5. 
R: SEDs for 7-band SWIRE-Lockman galaxies with z $>$ 4.
}
\end{figure*}

\subsection{Problematic sources with $\chi^2_{\rm ir} >$ 5}
First we show the SEDs of problematic sources with  $\chi^2_{\rm ir} >$ 5 in the automated 6-template fit, for sources with redshifts 
determined from 10 optical and 
near infrared photometric bands (Figs 5-8).  Most have log(S500/S24)$>$ 2 and z $<$ 1.  For over half the sources the introduction
of the cold cirrus template ($\psi$=0.1) solves the problem of the poor fit.  
In many cases none of our existing templates can match the observed SEDs.  Since it is implausible to postulate emission from cold
dust with a luminosity exceeding that of the illuminating starlight, the association of the submillimetre emission with
the chosen SWIRE galaxy must be incorrect.  A possible explanation is provided by galaxy lensing.  In several cases 
we have indicated fits to the submillimetre data with an Arp220 starburst model at a higher redshift than that of the SWIRE 
galaxy.  

\begin{figure*}
\includegraphics[width=7cm]{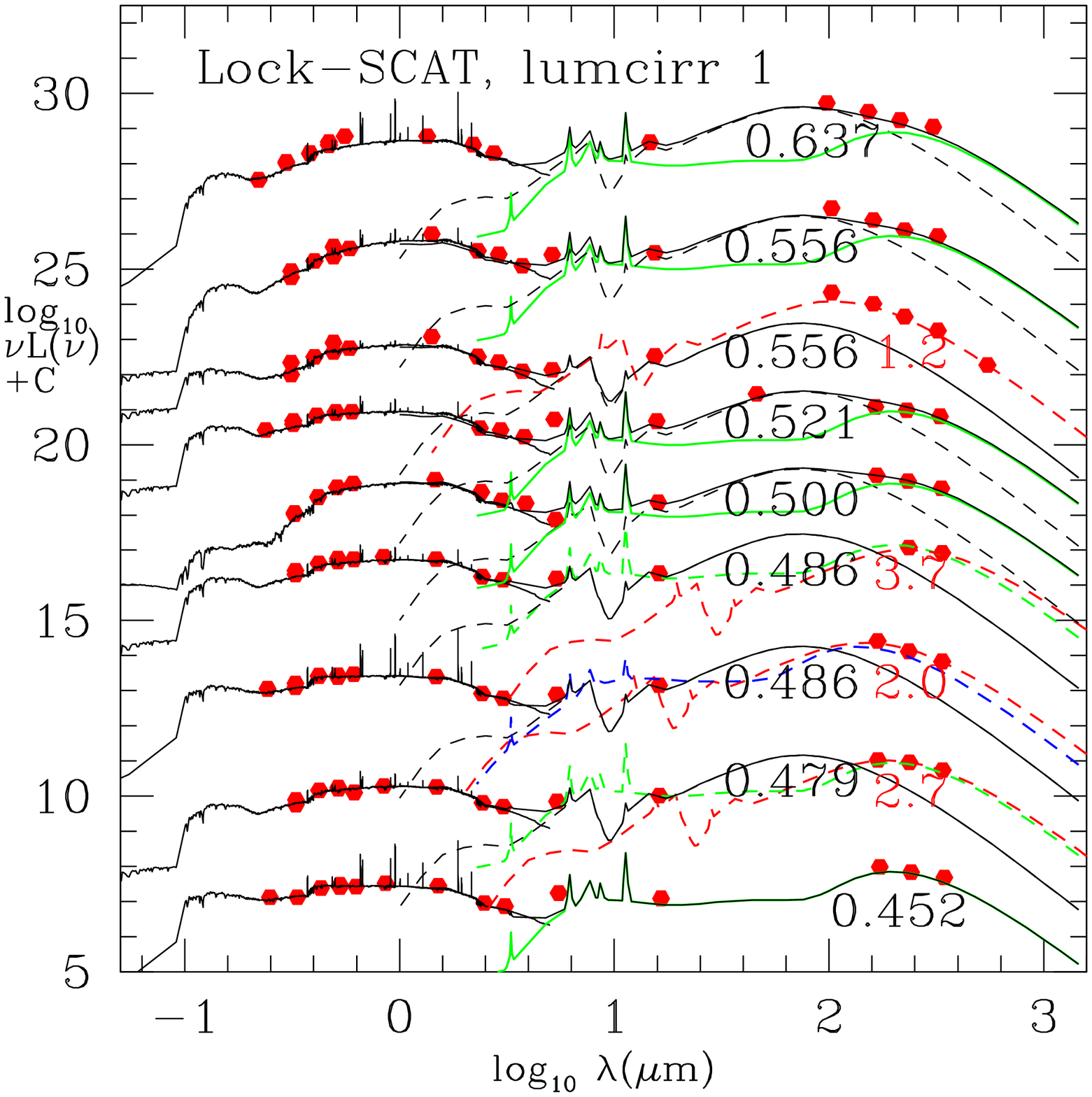}
\includegraphics[width=7cm]{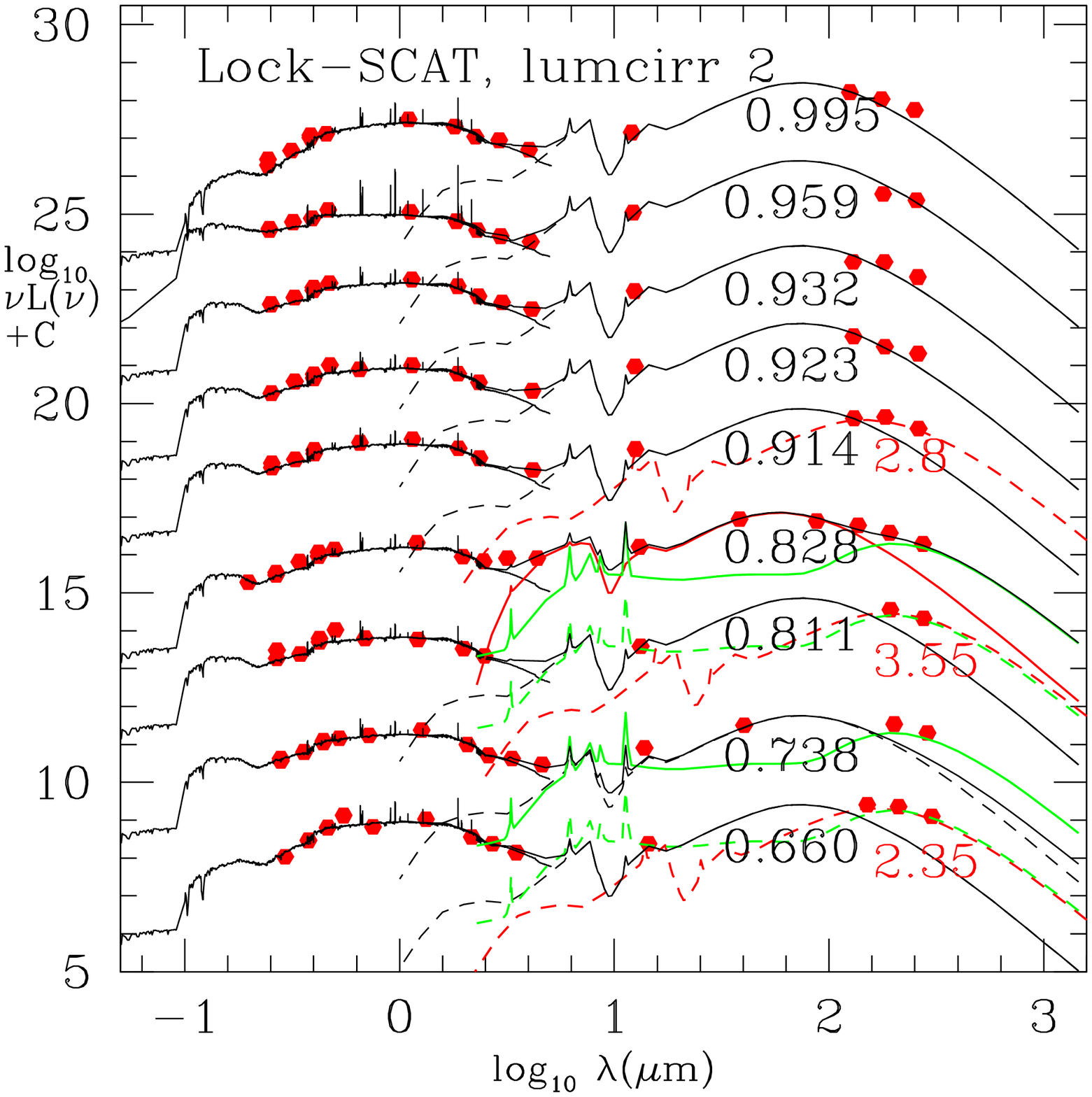}
\caption{SEDs for SWIRE-Lockman galaxies with $L_{\rm cirr} > L_{\rm opt}$.  Colour coding as in Fig. 5.
}
\end{figure*}

\subsection{Sources with good infrared template fits}
We now show the SEDs for 69 sources with redshifts determined from at least 12 optical-nir photometric bands and with $\chi^2_{\rm ir} <$ 5 (Figs 9-12).
In almost all cases the optical and near infrared data are well fitted by the galaxy or QSO templates and
the photometric redshifts look very credible.  The infrared template fits to $\lambda > 4.5 \mu$m data
also look good.  20 galaxies need a cold cirrus component and 3 are gravitational lens candidates. The
latter are cases where in the automatic fitting the submillimetre data are fitted with a cirrus component
with luminosity greater than that of the starlight, so a physically implausible fit despite the acceptable $\chi_{ir}^{2}$.

\begin{figure*}
\includegraphics[width=7cm]{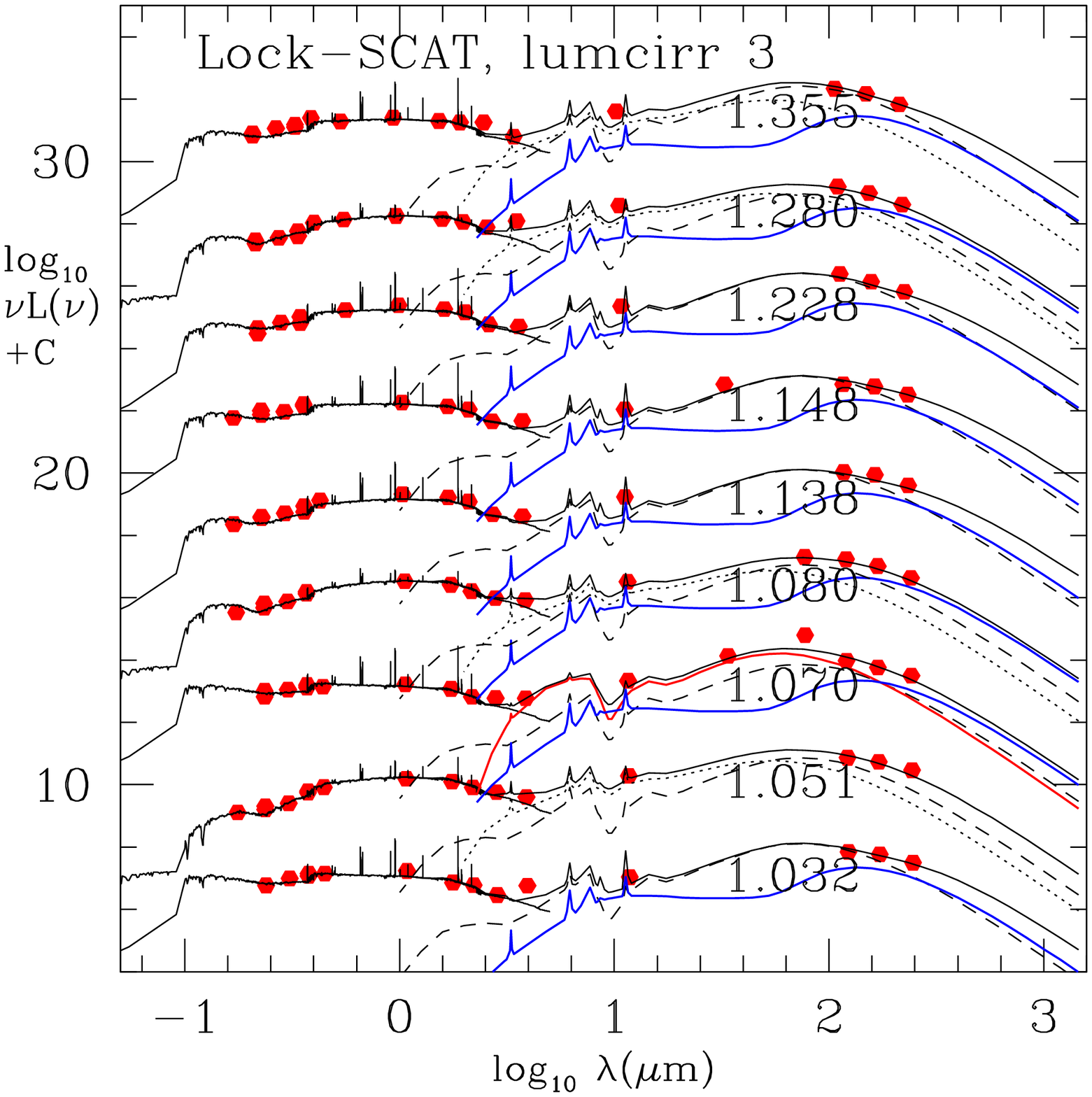}
\includegraphics[width=7cm]{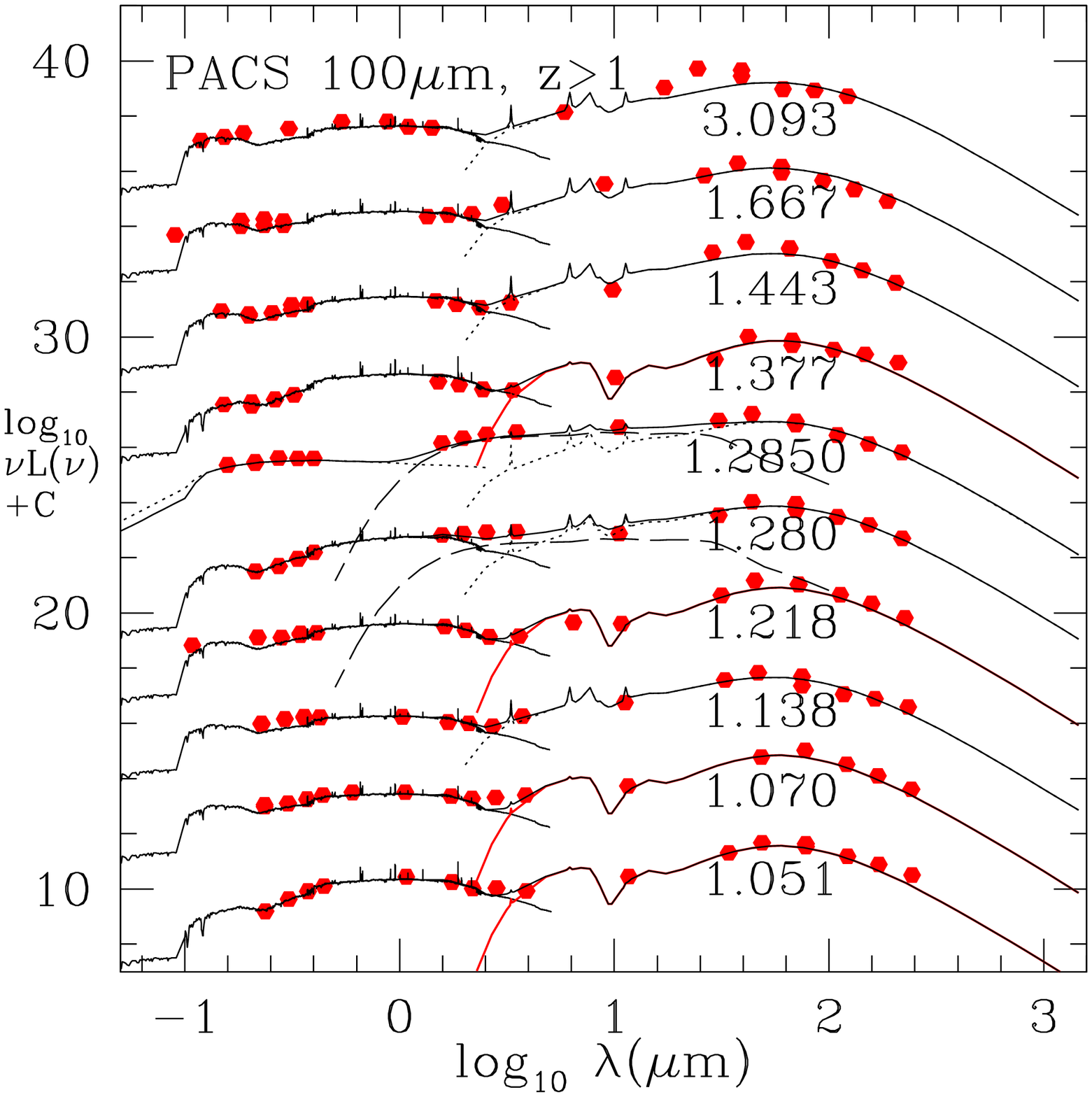}
\caption{LH: SEDs for SWIRE-Lockman galaxies with $L_{\rm cirr} > L_{\rm opt}$. 
RH: SEDs for SWIRE-Lockman galaxies with redshifts $>$ 1,determined from at least 7 photometric bands, and PACS 100 and 
160 $\mu$m detections.
}
\end{figure*}

The requirement of 12 optical and near infrared bands biases the sample against high redshifts, so we also show
the SEDs for sources with 7 photometric band redshifts, $\chi^2_{ir} <$ 5, and z $>$ 4 (Fig 12R).

We also incorporated photometry from the PACS data (Poglitsch et al. 2010).Ê The Lockman-SWIRE field was observed 
by HerMES (AOT Set no. 34, Oliver et al. 2012) using the SPIRE/PACS parallel mode.Ê Maps were reduced using the
Unimap software (Piazzo et al 2012).ÊÊ Photometry was estimated by constructing a "beam" 
from stacking the PACS map at the location of all 967 sources and fitting this beam to the location of each
source (this does not include a correction for the extended PACS beam).Ê We found
136 with S(100) $>$ 80 mJy, or S(160) $>$ 120 mJy. These thresholds (corresponding approximately
to 3 detections and >95 percent reliability) were selected after comparison
with SWIRE 70 and 160 $\mu$m fluxes, with predictions from our template fits, and from 
examination of individual SEDs.
When these 136 sources lie in our SED plots (Figs 5-14), the PACS 100 and 160 $\mu$m fluxes 
have been included and these are generally consistent with
the SWIRE 70 and 160 $\mu$m fluxes and the fitted models. We have also shown SEDs for 
10 sources with PACS 100 and 160 $\mu$m fluxes and z $>$ 1 (Fig 14R).  The PACS fluxes 
agree well with the fitted models.

\begin{figure}
\includegraphics[width=7cm]{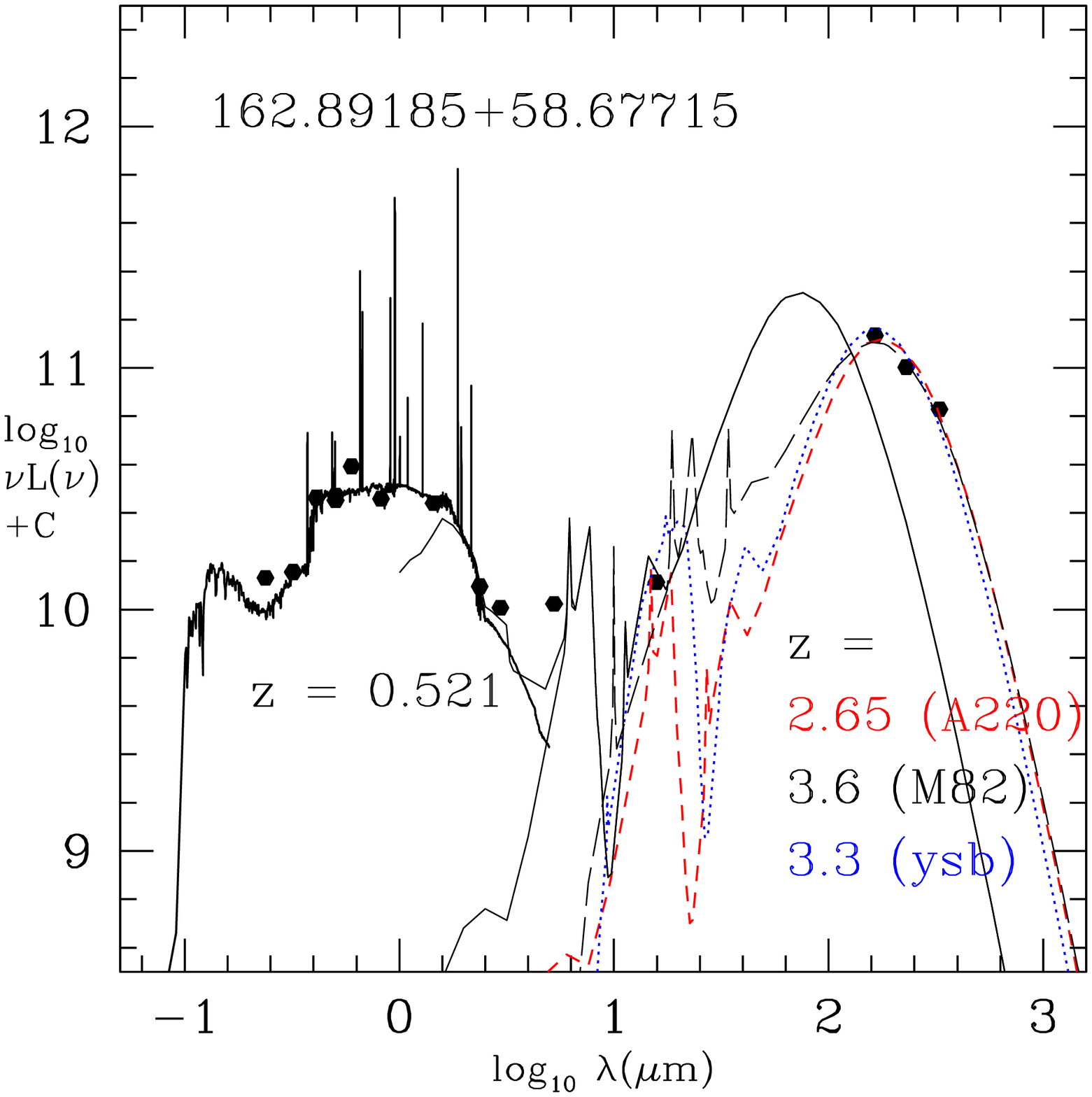}
\caption{SEDs for a candidate lensed galaxy (bottom galaxy in Fig 11R) with photometric redshift determined from 12 photometric bands, showing fits 
with Arp 220 (dashed), M82 (dash-dotted) and young starburst (dotted) templates.
}
\end{figure}

\subsection{Problematic sources with $L_{\rm cirr} > L_{\rm opt}$}
Figs 13,14L show SEDs for sources with at least 10 optical-nir photometric bands, and with $\chi^2_{ir} < 5$, but where $L_{cirr} > L_{opt}$ 
in the automated 6-template fits.  While some can be fitted
adequately by including a cold cirrus template without violating the $L_{cirr} \le L_{opt}$ requirement (and all at z$>$0.92), 
many require a lensing model to fit the submillimetre data.

In summary, the total number of candidate lenses from Figs 5-14 is 36.
Is there an alternative explanation to lensing for these galaxies with excessive submillimetre radiation?  Cold cirrus is not an option
for these lensing candidates because the luminosity in cold dust would exceed that in starlight.  Our association procedure
ensures that there is no other more likely 24$\mu$m association for the submillimetre source.  Any other association 
would have an even worse SED-fitting 
problem.  It is possible that in some cases we have a chance association of a z$\sim$0.5
SWIRE galaxy whose true submillimetre output is weak with a high redshift luminous submillimetre galaxy which is
undetected in the SWIRE survey.  We discuss in section 5 the probability for this to happen.  

The lens model SED fits have an ambiguity about which starburst template to use.  All fits shown use the Arp 220 template, but Fig 15
illustrates ambiguity in redshift for a lensed galaxy from using the M82 or young starburst templates.  The values of
$(1+z_{lens})$ that we are quoting need to be multiplied by 1.25 if a young starburst model is preferred or by 1.36 if an
M82-like starburst is preferred.

In Section 4 we use these 36 candidate lenses to define regions of colour-redshift space to identify a
further 73 lensing candidates from the sources with less than 10 optical-nir photometric bands.

\begin{figure*}
\includegraphics[width=14cm]{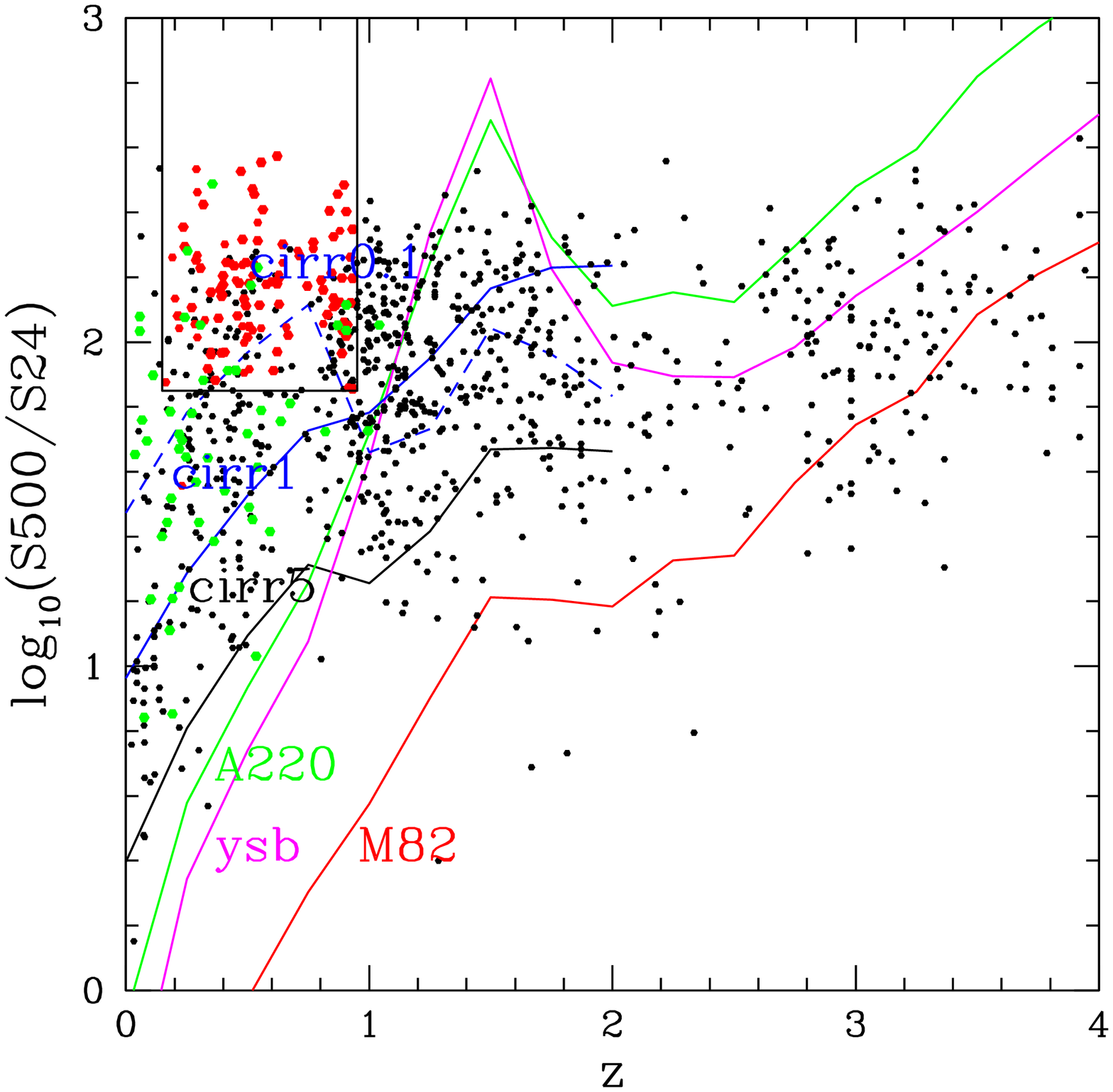}
\caption{500/24$\mu$m flux ratio versus redshift for SCAT Lockman sources with 5-$\sigma$ detections at 350 and 500$\mu$m,
and which are also associated with SWIRE 24$\mu$m sources (filled black circles). Filled red circles: 109 lensing candidates; green filled circles: 50 galaxies requiring cold dust template ($\psi = 0.1$).
}
\end{figure*}

\begin{figure*}
\includegraphics[width=7cm]{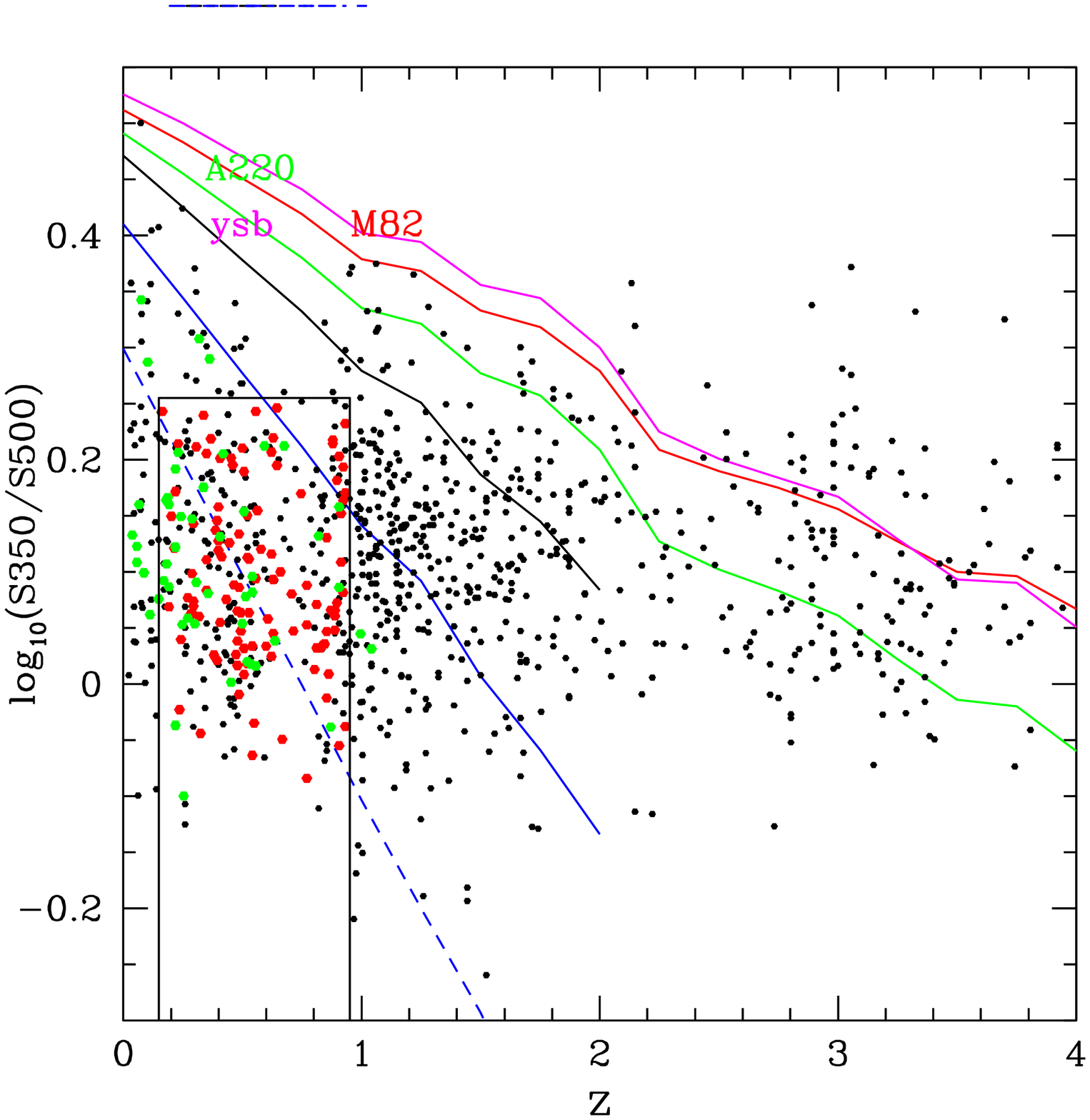}
\includegraphics[width=7cm]{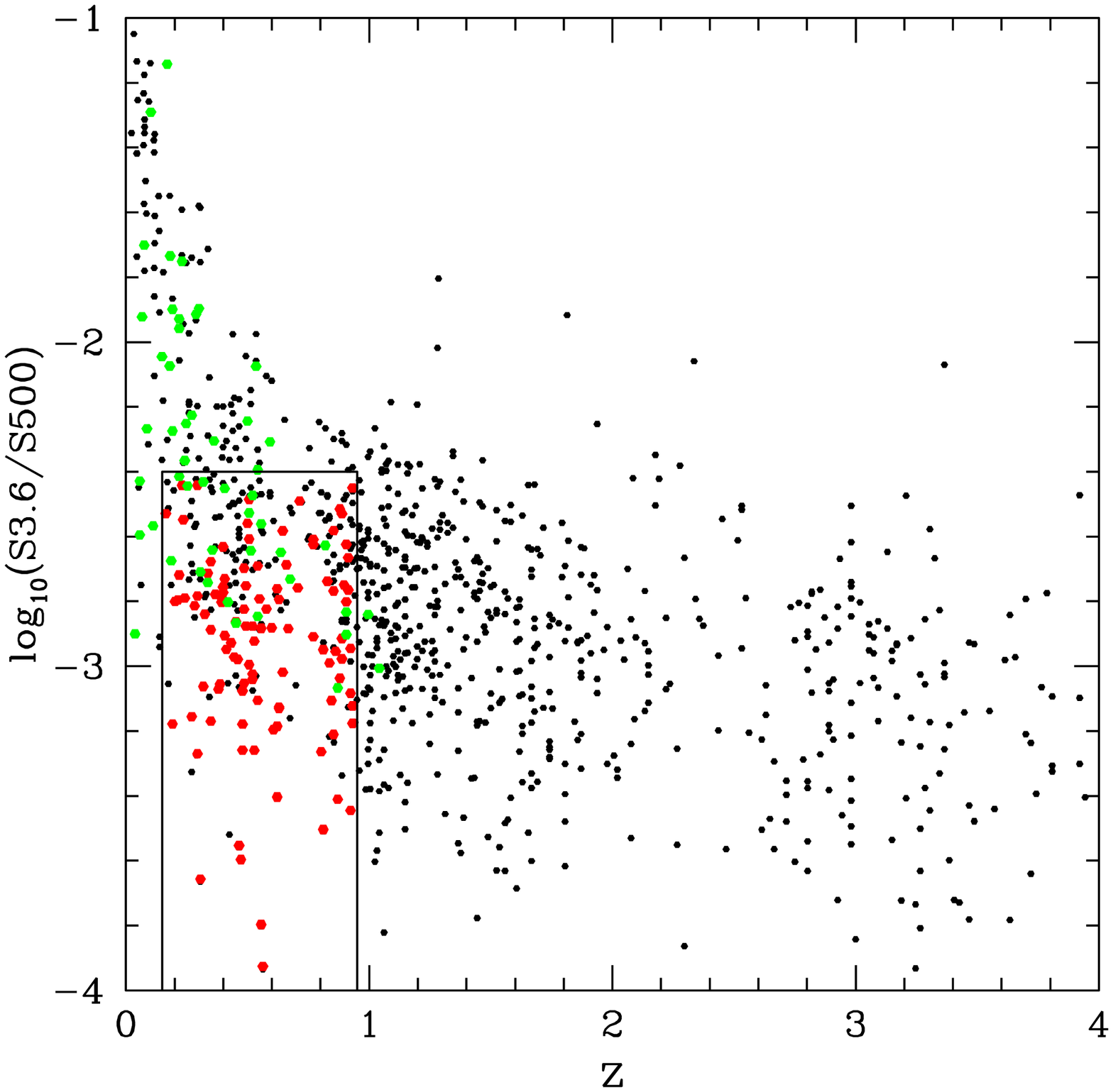}
\caption{L: 350-500$\mu$m colour versus redshift for SCAT Lockman sources with 5-$\sigma$ detections at 350 and 500$\mu$m,
and which are also associated with SWIRE 24$\mu$m sources.  Lensed sources shown in red, galaxies requiring cold dust
in green.
R: $3.6/500 \mu$m flux-ratio versus redshift for SCAT Lockman sources with 5-$\sigma$ detections 350 and 500$\mu$m
and which are also associated with SWIRE 24$\mu$m sources.
}
\end{figure*}

\section{Using colour-redshift diagrams to select lensing candidates}

Our SED modelling of problematic sources with 10-band photometric redshifts has identified 
36 candidate lensed galaxies, in which the optical and near infrared data define the lensing galaxy
and the submillimetre data are due to the background lensed galaxy.  We now explore whether we
can use colour-redshift plots to characterise lensing candidates amongst fainter sources, which generally 
have fewer photometric bands available.

The 36 galaxies occupy a rather well defined area in the plot
of $log_{10} S500/S24$ versus redshift (Fig 16), which can not be reached by our infrared templates.
In addition  Wardlow et al (2013) have shown that gravitational lens candidates have well-defined 250-350-500$\mu$m
colours.  Fig 17L shows $log_{10} S500/S350$ versus redshift
for SCAT Lockman sources, with lens candidates shown in red.  The Wardlow et al criterion, S350$/$S500 $<$ 1.8
does indeed include all our lensing candidates selected on the basis of their SEDs.  We do not try to use the 250$\mu$m
flux since some of our sources are not detected at 250$\mu$m.

\begin{figure*}
\includegraphics[width=7cm]{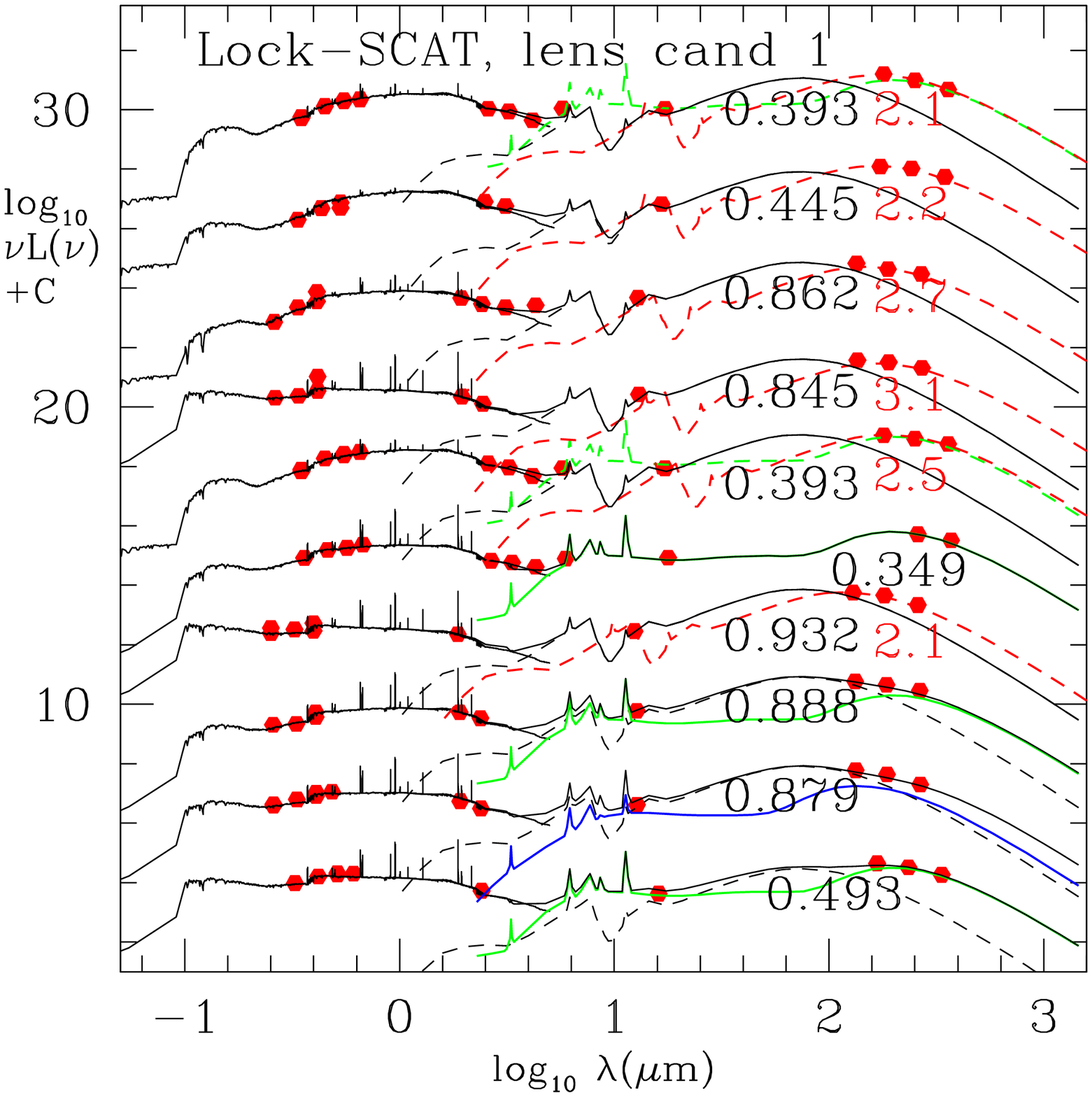}
\includegraphics[width=7cm]{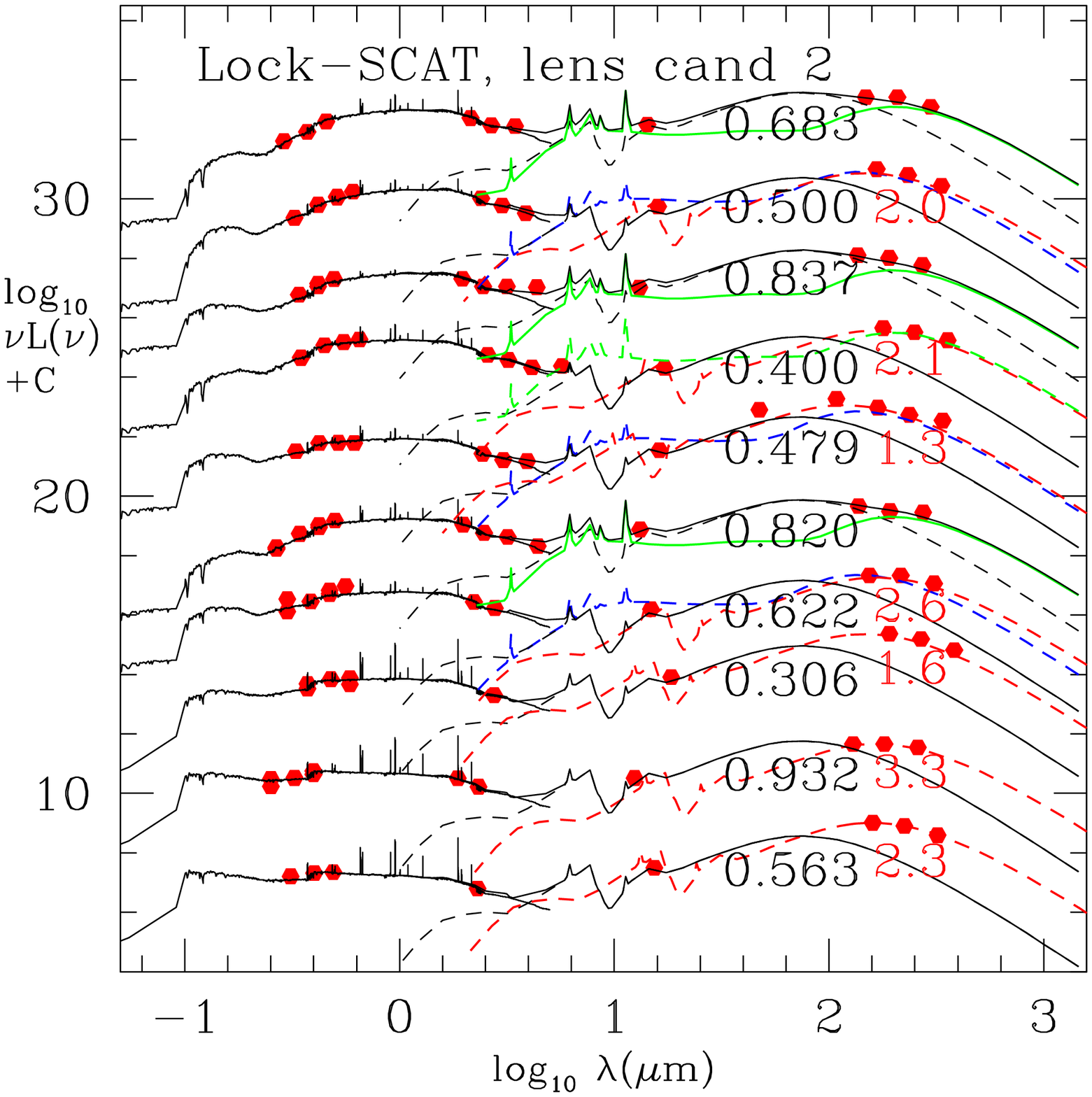}
\caption{SEDs for  SWIRE-Lockman galaxies with $<$ 10 photometric bands which are lens candidates
according to the colour-redshift criteria of section 4. 
}
\end{figure*}

\begin{figure*}
\includegraphics[width=7cm]{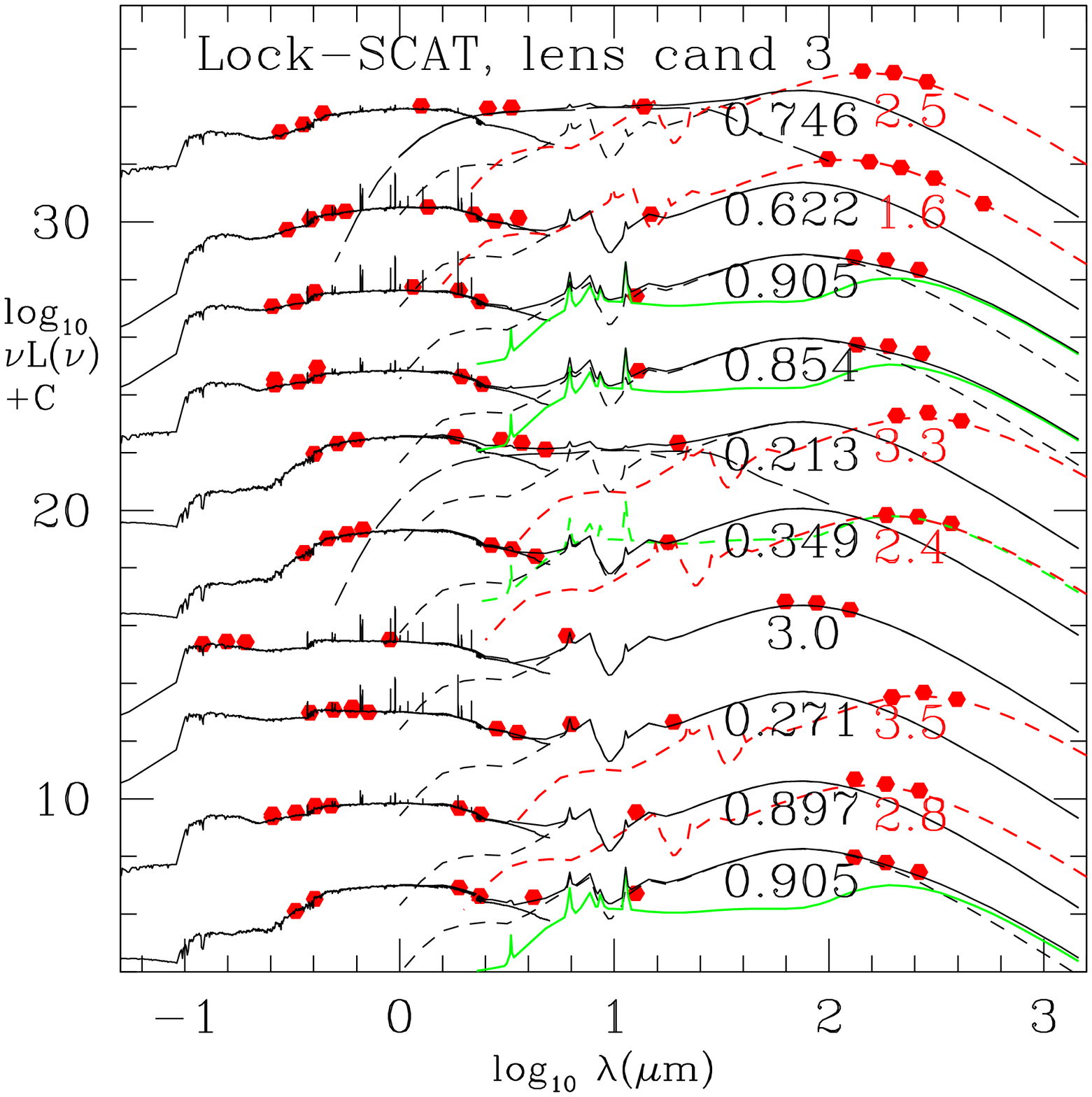}
\includegraphics[width=7cm]{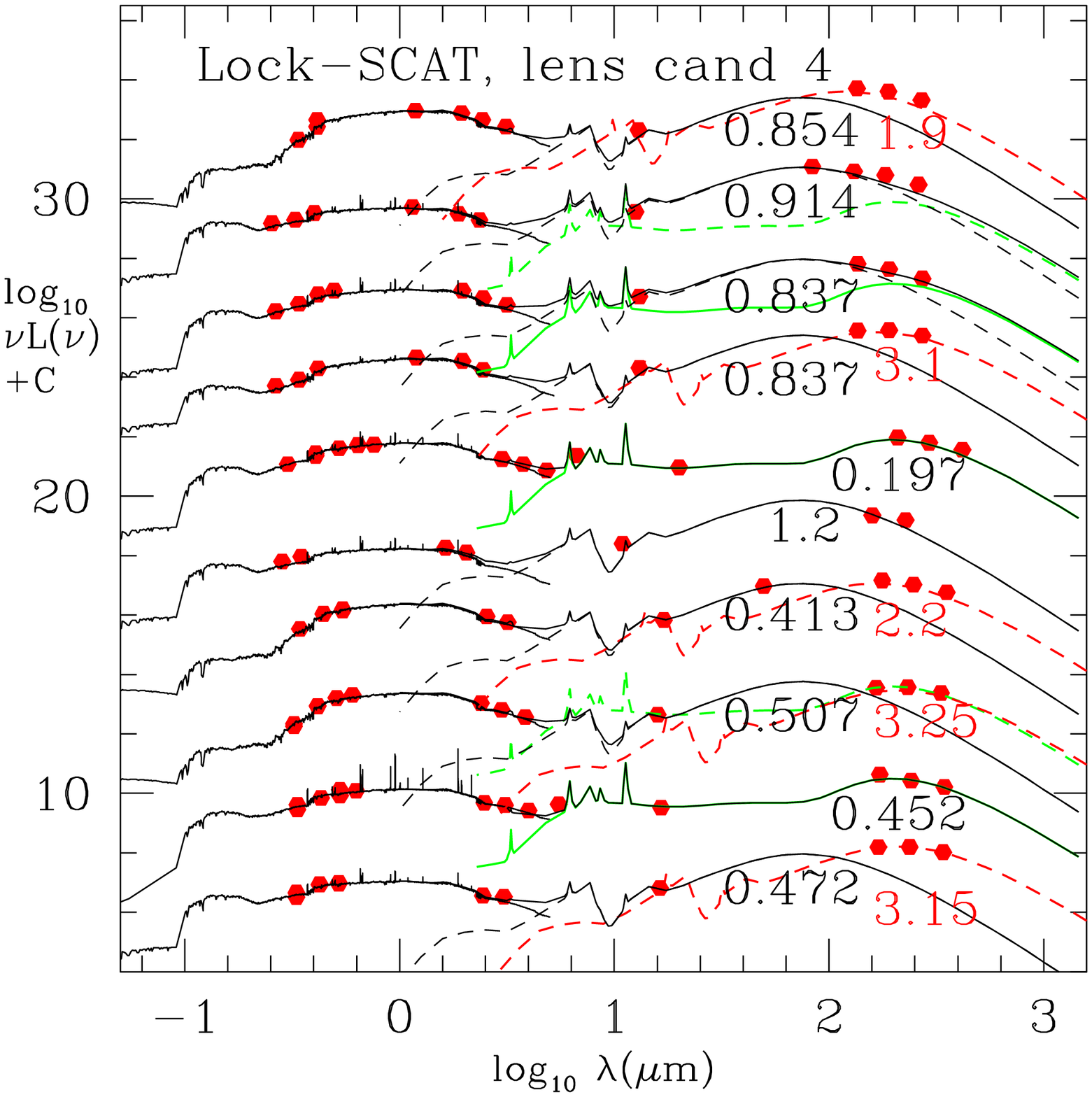}
\caption{SEDs for  SWIRE-Lockman galaxies with $<$ 10 photometric bands which are lens candidates
according to the colour-redshift criteria of section 4.
}
\end{figure*}

To capture the $L_{\rm cirr} > L_{\rm opt}$ indication of lensing we have plotted $log_{10} S3.6/S500$ versus redshift (Fig 17R),
with lensing candidates on the basis of their SEDs in red.  Again the lensing candidates occupy a well-defined area of this
diagram.

We can now define criteria which lensed candidates in our sample are likely to satisfy:  0.15 $<$ z $<$ 0.95, $log_{10} (S3.6/S500) < 0.6$, 
$S350/S500 < 1.8$, and $log_{10} (S500/S24) > 1.85$.  
Here we are assuming that the redshift determined from optical and near infrared photometry applies to the lensing galaxy.
We have modelled the SEDs of all the 117 galaxies satisfying all three criteria with redshifts based on less 
than 10 photometric bands (Figs 18-23).  Of these, 44 could be fitted with our standard 7 templates and without violating $L_{\rm cirr} \le L_{\rm opt}$ (21 of these involve cold dust).  The
remaining 73 are lensing candidates.  Of the 117 sources satisfying our colour constraints, 63$\%$ are lensing candidates while 18$\%$
require cold dust.  In total we have 109 lensing candidates, based on their SEDs, out of our total sample
of 967 Lockman-SCAT-SWIRE sample (11.3$\%$), or 8$\%$ of our Lockman-SCAT 500+350$\mu$m sample.  These include the Wardlow et al (2013) lensing candidate HLock06, 
which falls in the SWIRE area.  HLock01 is not in our catalogue but would satisfy our lensing criteria. The stellar masses of our candidate lensing galaxies mostly lie in the range $log_{10} M_*/M_{\odot}$ = 10.5-11.5, with a mean value 11.0.

Our lensing criteria could be on the conservative side, because in some cases submillimetre emission could be lensed even though there is an acceptable fit to the SED from a non-lensed model.  Some of the lensing
candidates will be chance associations of a high redshift submillimetre galaxy with a lower redshift 24 $\mu$m galaxy
and this can only be resolved by submillimetre intereferometric imaging.

\begin{figure*}
\includegraphics[width=7cm]{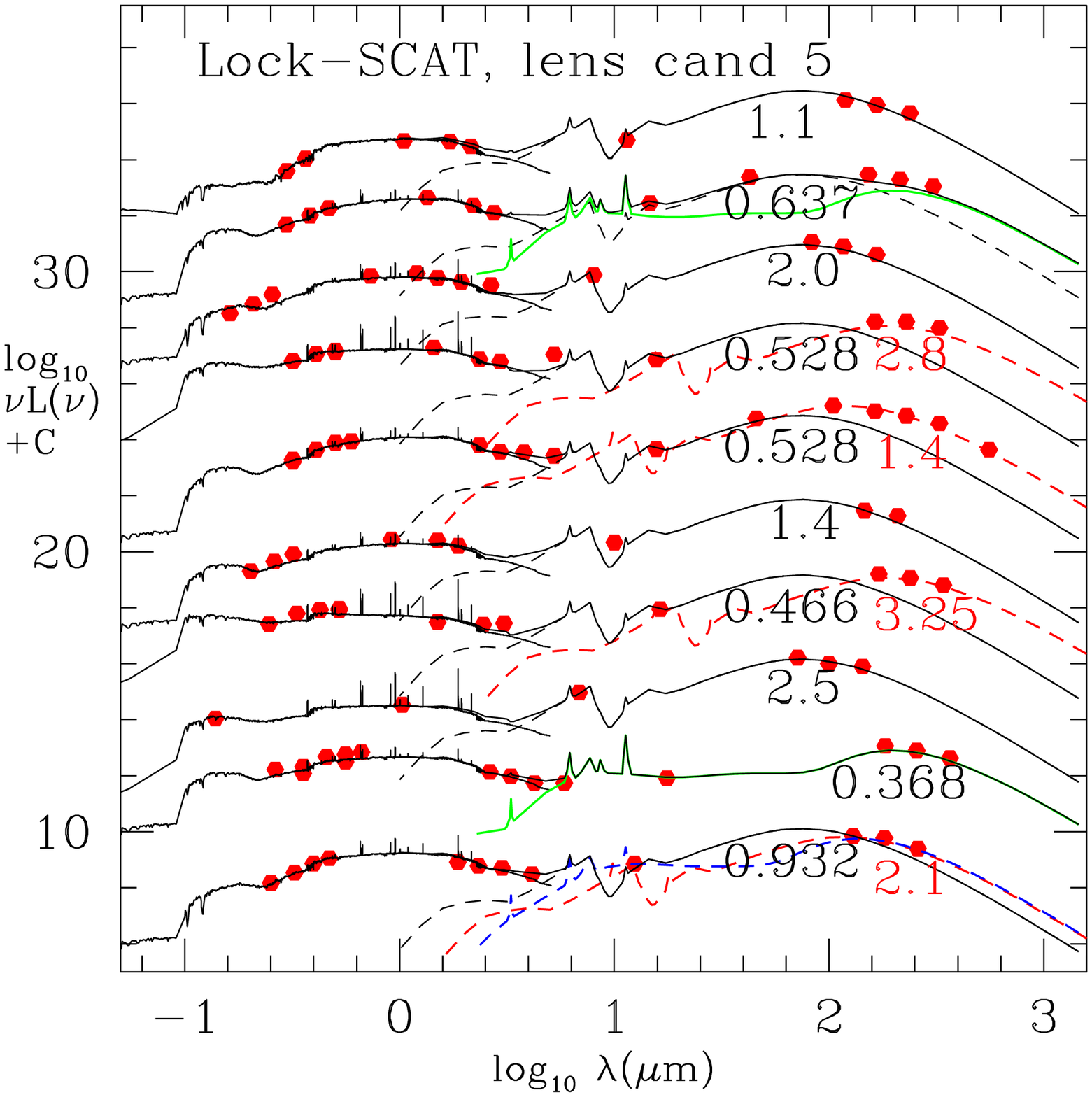}
\includegraphics[width=7cm]{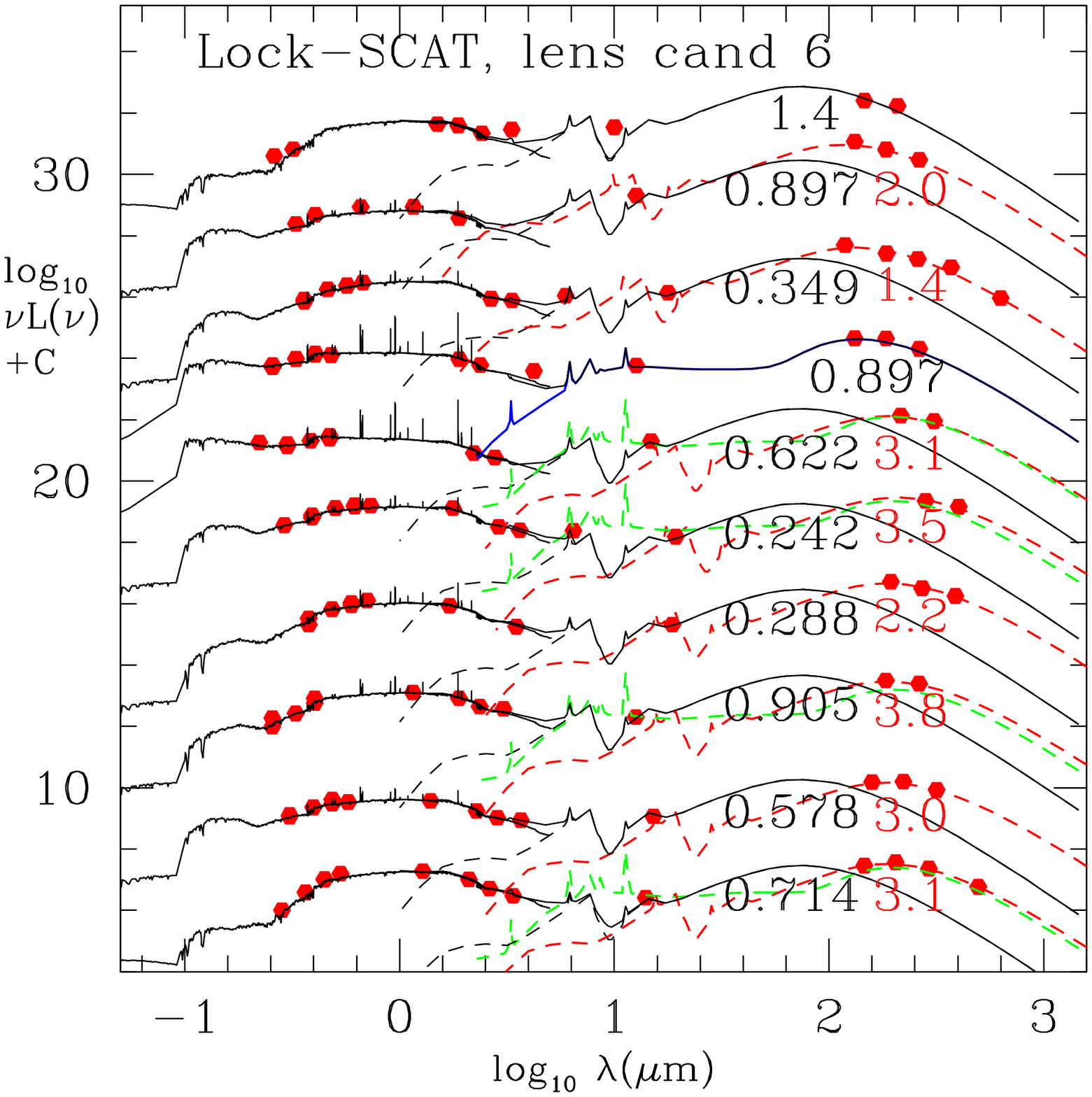}
\caption{SEDs for  SWIRE-Lockman galaxies with $<$ 10 optical-nir photometric bands which are lens candidates
according to the colour-redshift criteria of section 4.  Colour coding as in Fig. 5.
}
\end{figure*}

\begin{figure*}
\includegraphics[width=7cm]{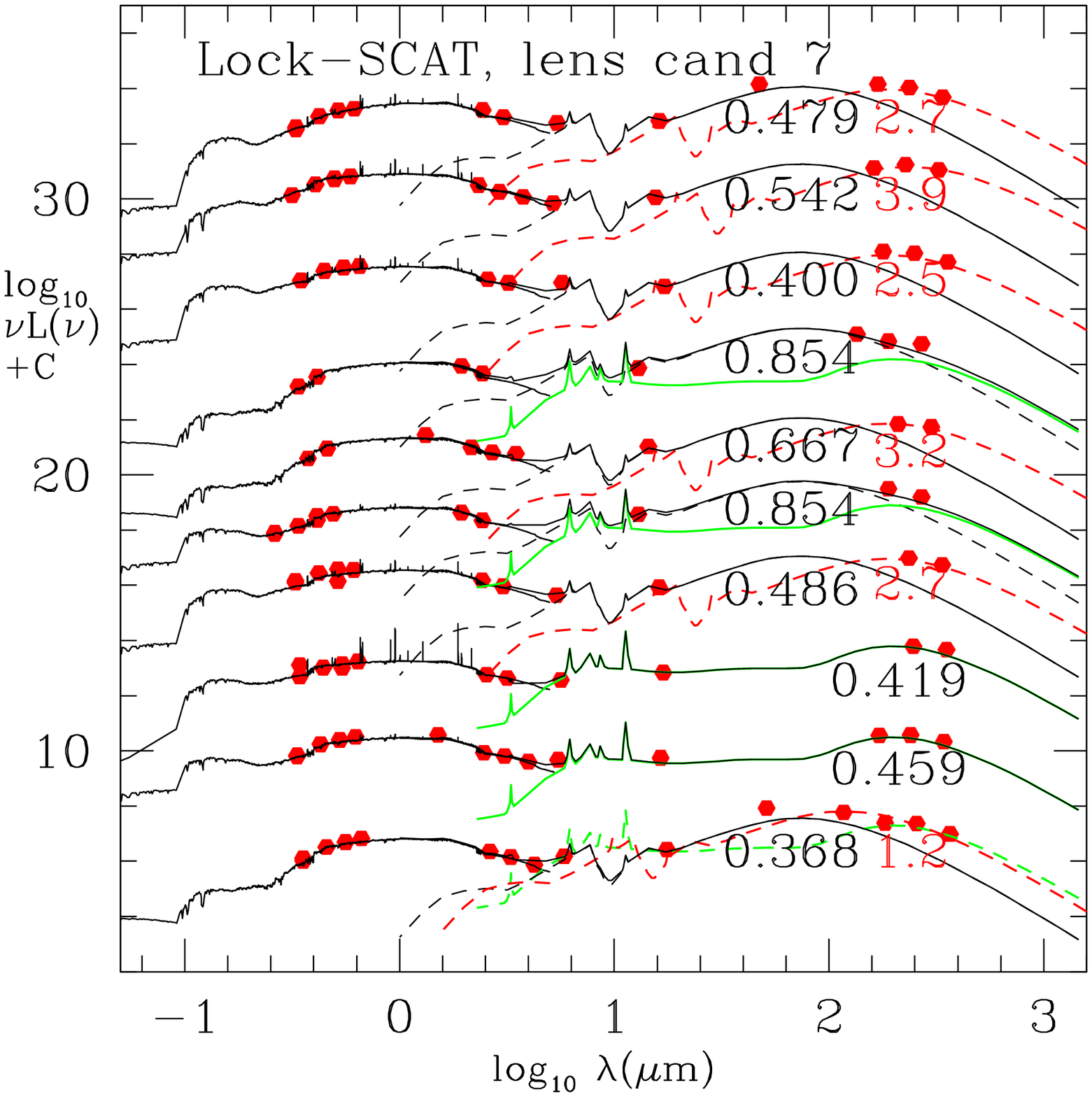}
\includegraphics[width=7cm]{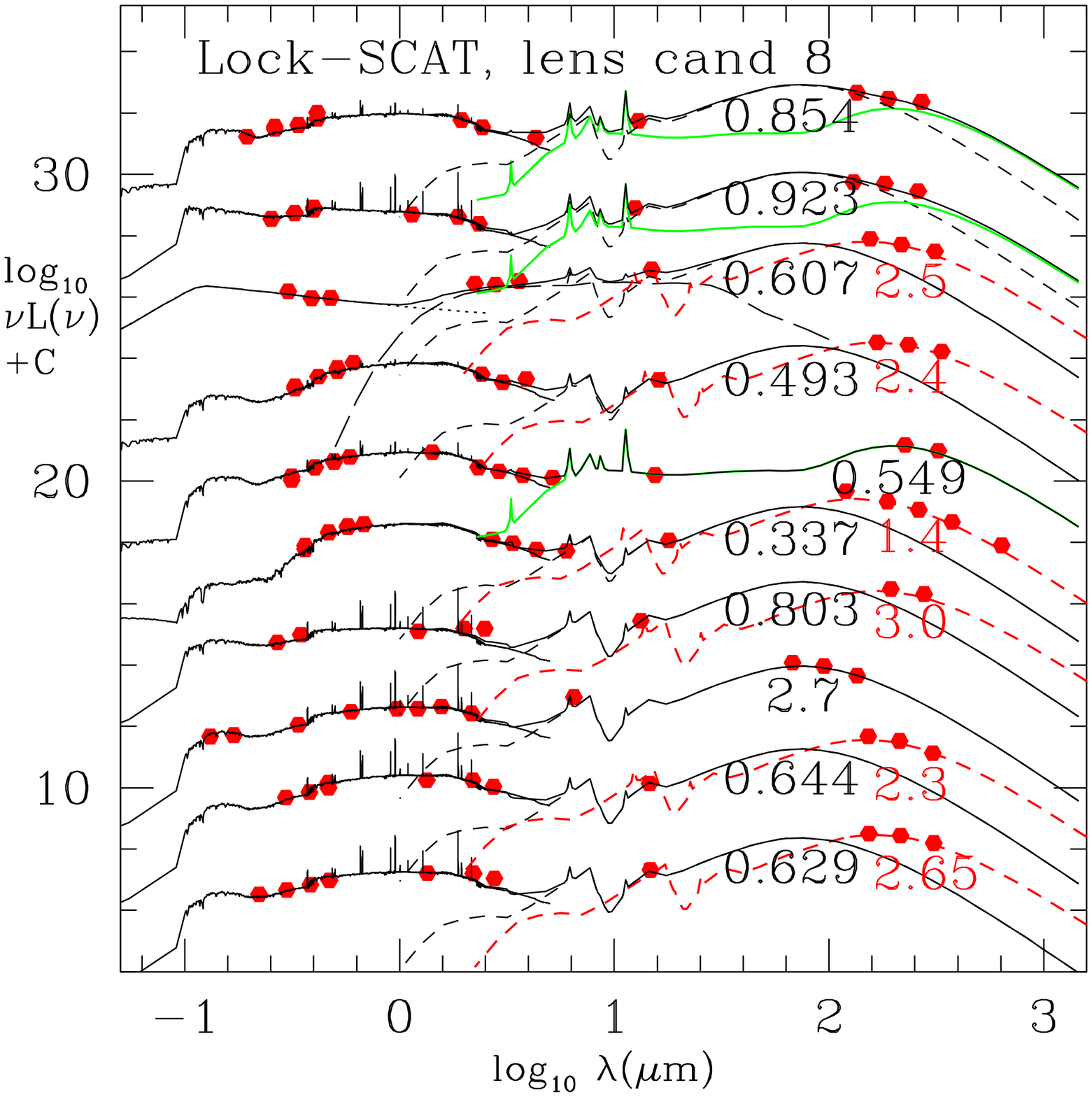}
\caption{SEDs for  SWIRE-Lockman galaxies with $<$ 10 photometric bands which are lens candidates
according to the colour-redshift criteria of section 4.
}
\end{figure*}

\begin{figure*}
\includegraphics[width=7cm]{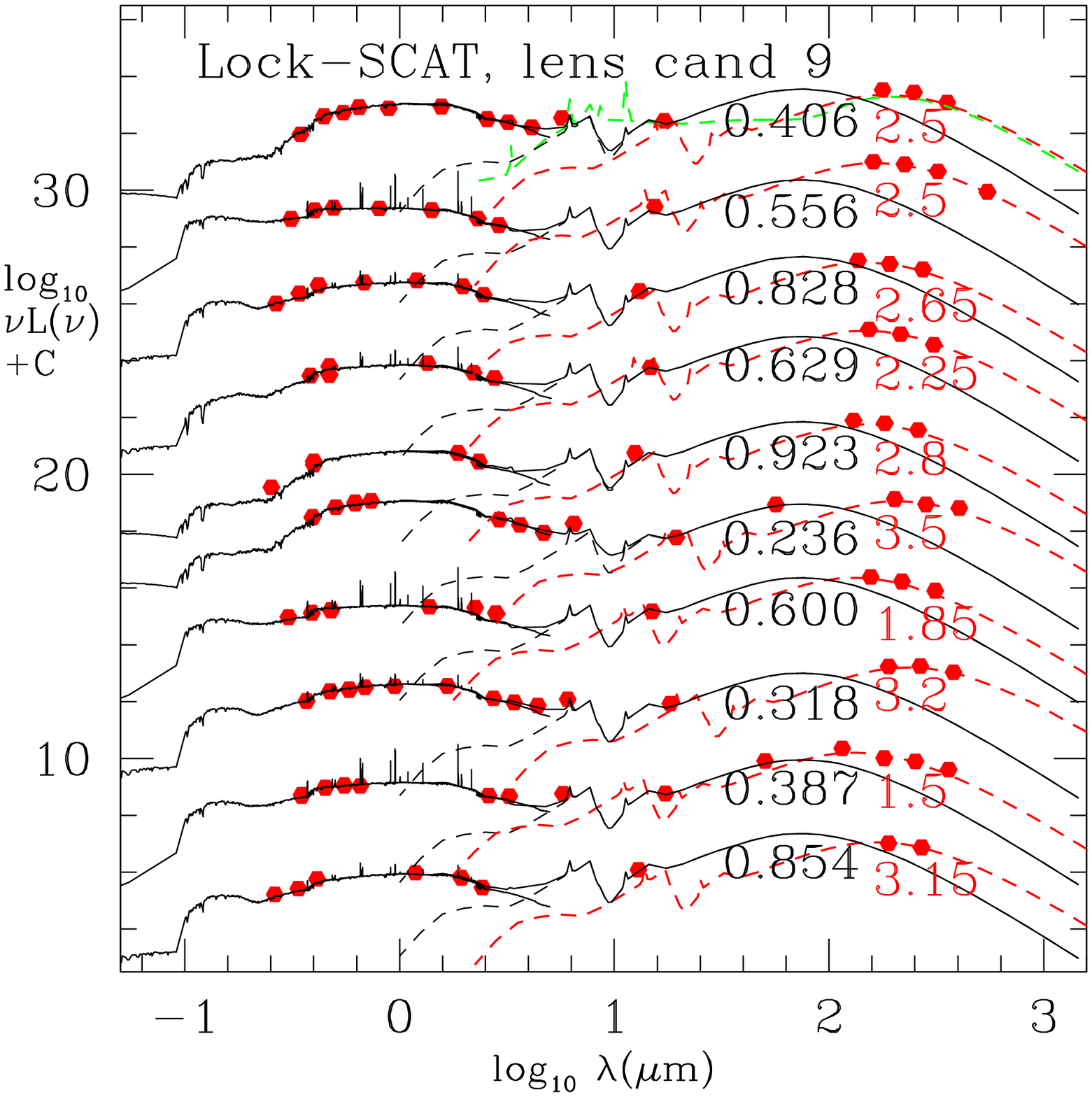}
\includegraphics[width=7cm]{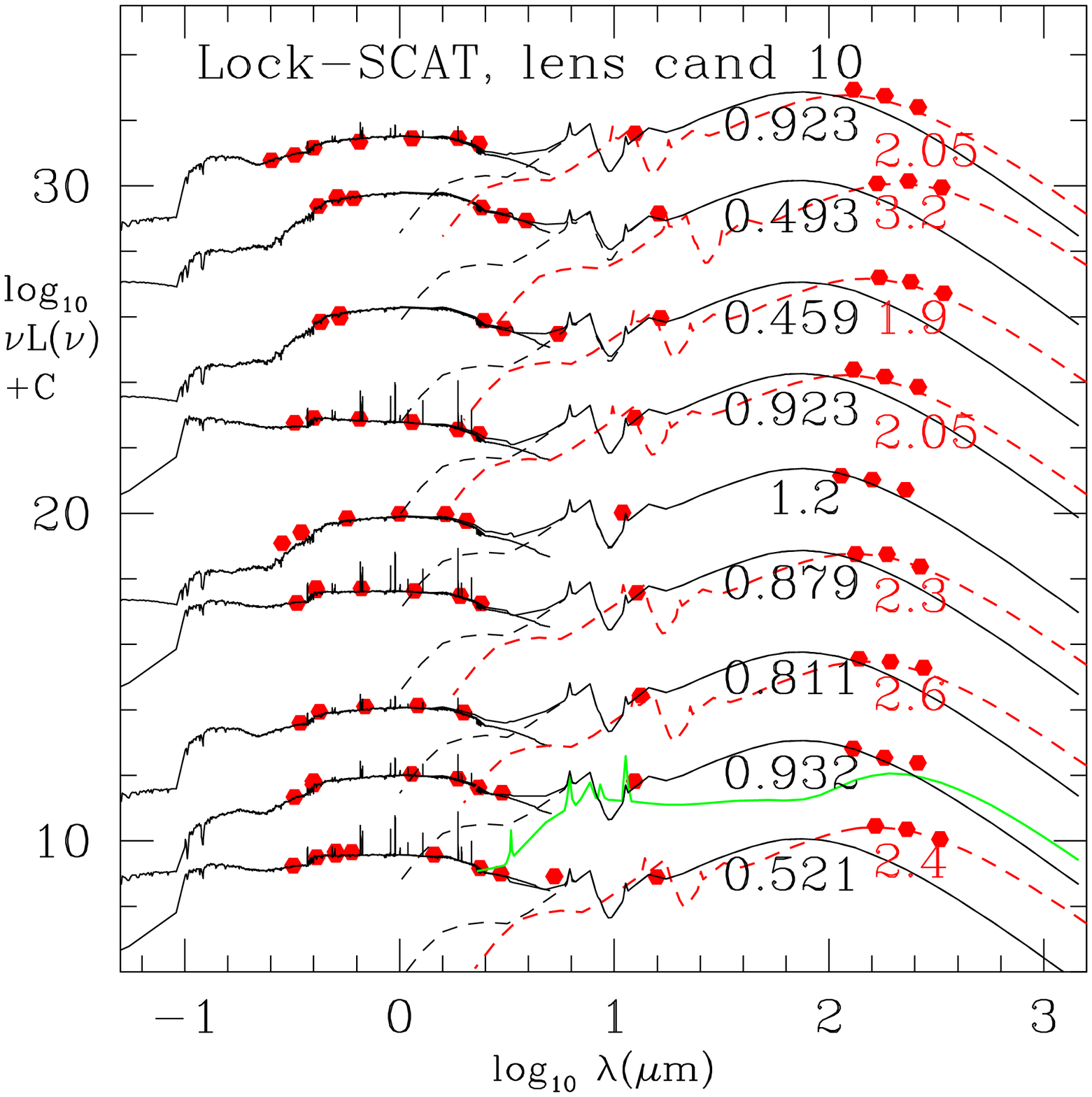}
\caption{SEDs for  SWIRE-Lockman galaxies with $<$ 10 photometric bands which are lens candidates
according to the colour-redshift criteria of section 4.
}
\end{figure*}

\begin{figure*}
\includegraphics[width=7cm]{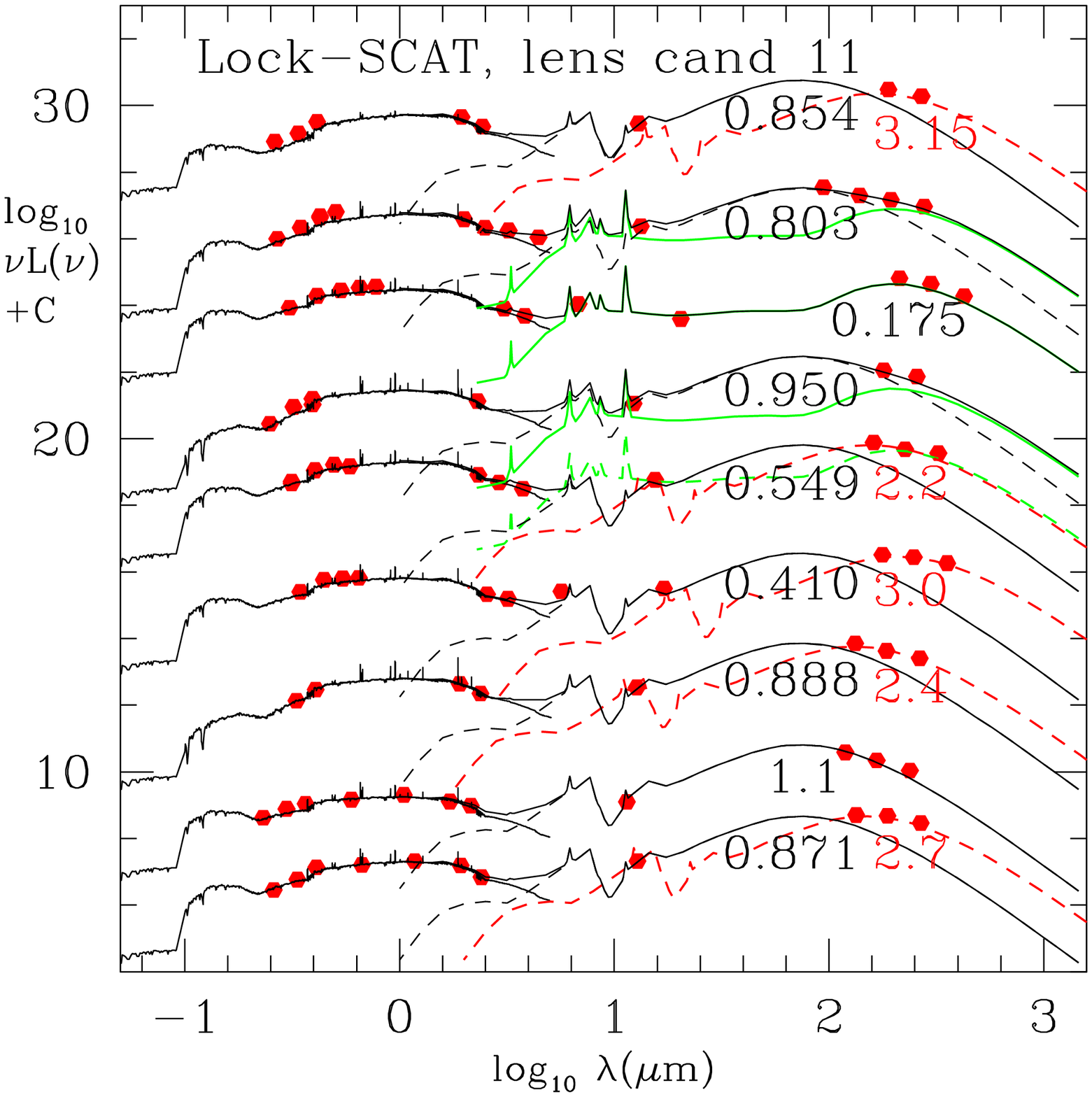}
\includegraphics[width=7cm]{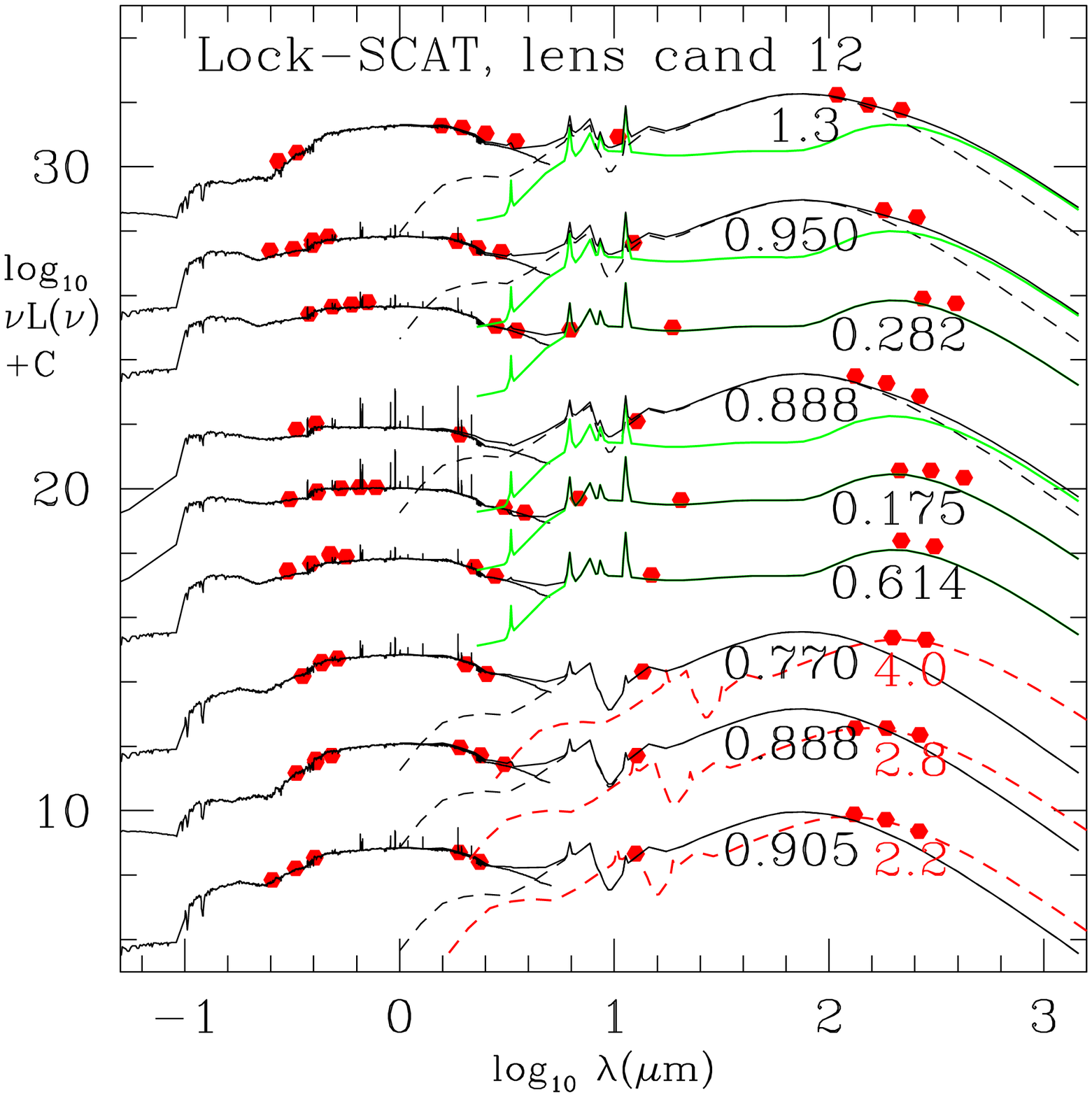}
\caption{SEDs for  SWIRE-Lockman galaxies with $<$ 10 photometric bands which are lens candidates
according to the colour-redshift criteria of section 4.
}
\end{figure*}

\section{Confusion and chance associations}
The restriction to 5-$\sigma$ detections should ensure that most of our sources are distinct independent galaxies.  The source-density
of our sample corresponds to one source per 36 beams, so the probability of a source being a blend of two fainter sources in the
beam is low, ignoring correlations.  However
the limited {\it Herschel} resolution, combined with the high source-density of {\it Spitzer} 24$\mu$m sources, means
that the issue of incorrect association between SPIRE and SWIRE sources is a serious one.  This is especially important
when we are claiming to identify the presence of unusually cold dust or of gravitational lenses from anomalous
spectral energy distributions.

For our 109 gravitational lens candidates and the 50 sources whose SEDs we have claimed require the presence of cold dust
we have therefore very carefully examined the regions around the SPIRE sources for alternative associations.  We have looked 
for all additional 24$\mu$m associations (with S24 $>$ 100 mJy) within 20 arcsec of the SPIRE position, which are either
closer than our preferred association or are brighter at 24$\mu$m (or both).   22 of the 109 lens candidates yielded a total of
25 alternative associations, while 8 of the 50 galaxies requiring cold dust yielded a total of 10 alternative associations. The
remaining 87 lens candidates and 42 cold dust galaxies did not have an alternative 24 $\mu$ association.  

We have modelled the SEDs of these 35 sources, assuming the 250-500$\mu$m flux belongs to these alternative
associations rather than our preferred association based on the predicted 450$\mu$m flux.  Of the 25 lens candidates alternative
associations, 20 still needed a lensing model, 5 did not.   Thus for 95$\%$ of our lens candidates, there is no alternative to a 
lensing model, though there may be ambiguity about which galaxy is the lens in some cases.  For $\sim 5\%$ of our lens candidates, 
the choice of a lensing model may be due to the wrong association.   Clearly there is a need for submillimetre interferometry
to confirm the reality of these lensing candidates, and to clarify which 24$\mu$m source is the lens in the ambiguous cases.

Of the 10 cold dust galaxies alternative associations, 6 require a lensing model, one requires a cold dust component, and 3 do not
require a lensing model or cold dust component.  We see from Figs 16 and 17 that many of the cold dust galaxies occupy the
same regions of the colour-redshift diagrams as the lensing candidates, so there is ambiguity in the interpretation of their
anomalous SEDs.  If the luminosity in the submillimetre component is less than the optical-nir starlight luminosity, we have
assumed a cold dust interpretation, but a low-luminosity lensed galaxy interpretation cannot be excluded.  Again
submillimetre interferometry is the key to confirming the correctness of the cold dust interpretation, since these would
be expected to be unusually extended because $\psi$ is a measure of surface brightness, so for a given luminosity lower 
$\psi$ implies a larger diameter for the dust cloud.

We also need to estimate the probability that the association of a z = 0.15-0.95 SWIRE galaxy with a 500 $\mu$m source is spurious
because the submillimetre source is in fact a background high redshift galaxy without a SWIRE detection, which is associated
with a foreground SWIRE galaxy by chance.  
For each of the 368 500 $\mu$m sources which did not find a 24 $\mu$m association, which are our candidate background high redshift
galaxies, we estimate the probability of a chance association with a
 z = 0.15-0.95 SWIRE 24 $\mu$m galaxy, requiring that our lensing criteria (section 4) be satisfied.  We estimate 24 chance
associations within the 7.5 degree Lockman-Hermes area, so $\sim 22 \%$ of our lensing candidates could be chance associations
of a 24 $\mu$m galaxy with a background high-z 500 $\mu$m galaxy.

Finally we should consider whether the attribution of sources as lensing candidates could be due to catastrophic outliers
in the photometric redshift estimate.  Of the 109 lensing candidates, 66 have redshifts determined from at least 9 photometric
redshift bands, for which the probability of a catastrophic outlier is $< 1\%$ (Rowan-Robinson et al 2013); 35 have redshifts
determined from 5-8 bands, for which the probability of a catastrophic outlier is 1-3$\%$, 8 have redshifts determined from 3-4
bands, for which the probability of a catastrophic outlier is 10-20$\%$; and 2 have redshifts determined from 2 bands, which are highly uncertain.  So catastrophic outliers could account for up to 5 of the lensing candidates.  From the SEDs all the photometric redshifts
look plausible. Two are confirmed by spectroscopic redshifts.

To summarise, we estimate that $\sim 5\%$ of the candidate lensed galaxies could be due to catastrophic photometric redshift outliers,
$\sim 5\%$ could be cases where the wrong SWIRE association has been chosen, and $\sim 20\%$
are likely to be chance associations of a high redshift 500$\mu$m galaxy with a lower redshift
24$\mu$m galaxy.

\section{Discussion}

Figure 24L shows the redshift distribution for the 858 unlensed SCAT 500$\mu$m sources in Lockman which are associated with
SWIRE galaxies (or QSOs).  The 368 unassociated SCAT sources have been shown, arbitrarily, distributed uniformly between redshift
1.5 and 5.  Fig 24R shows the redshift distribution for the 109 candidate lenses (in blue) and for the candidate lensed galaxies (in red)
based on an Arp220 fit to the submillimetre data.

\begin{figure*}
\includegraphics[width=7cm]{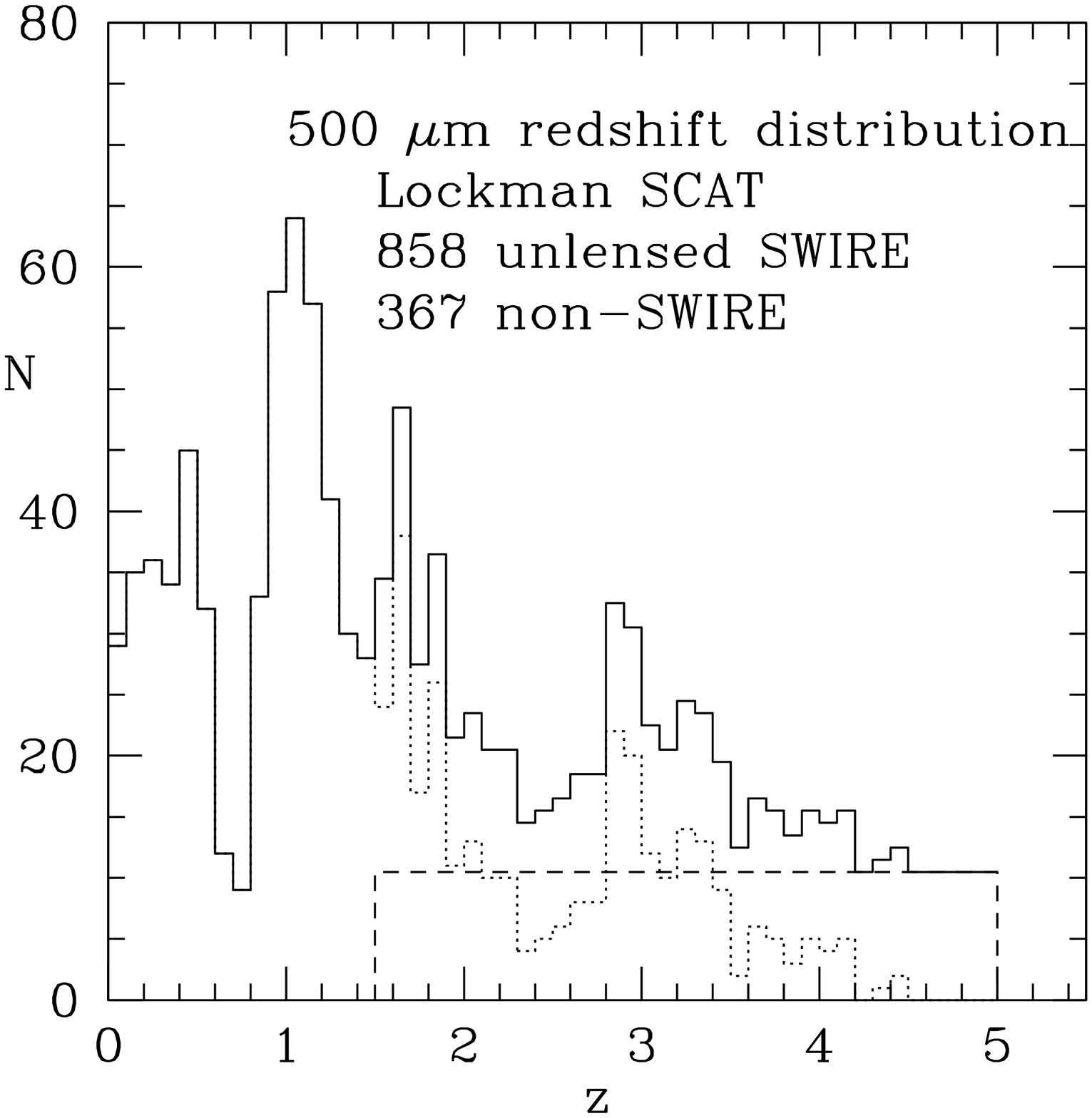}
\includegraphics[width=7cm]{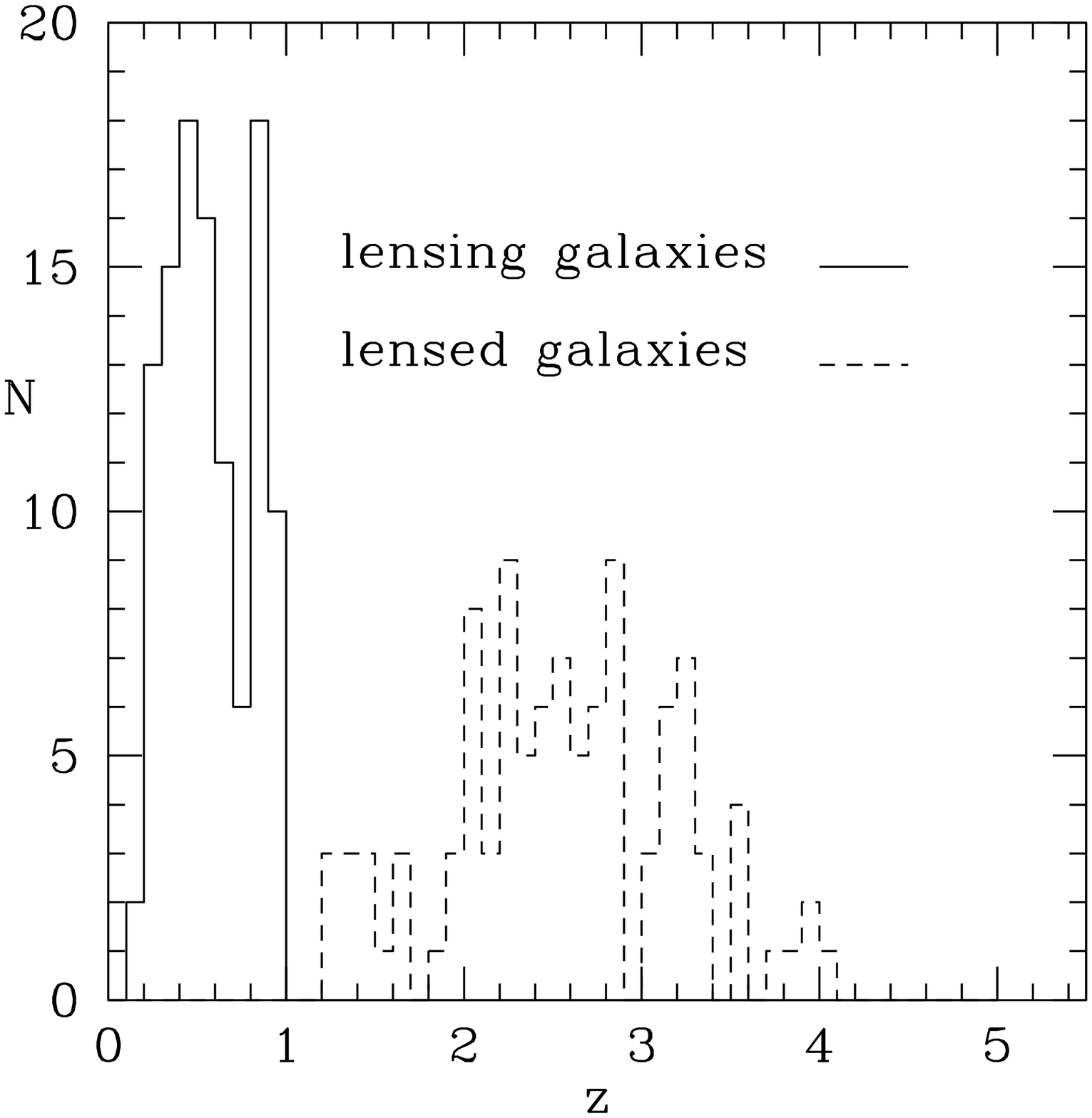}
\caption{L: Redshift distribution for unlensed SCAT 500$\mu$m sources in Lockman.  Sources not associated with
SWIRE galaxies have been shown uniformly distributed between z = 1.5 and z = 5.
R: Redshift distribution for SCAT 500$\mu$m lensing galaxies (solid locus) and for corresponding lensed galaxies (dashed locus).
}
\end{figure*}

Figure 25L shows the distribution of infrared luminosity (1-1000$\mu$m), $L_{\rm ir}$, with redshift for the 858 unlensed SCAT 500$\mu$m 
sources in Lockman which are associated with SWIRE galaxies (or QSOs), colour-coded with the dominant contribution to
the luminosity.  The absence of sources to the lower right reflects the selection effects at 24, 350 and 500
$\mu$m, which are different for each template type.  
598 galaxies (70$\%$) are ultraluminous ($L_{\rm ir}>10^{12} L_{\odot}$)
and 225 (26$\%$) are hyperluminous ($L_{\rm ir}>10^{13} L_{\odot}$).  295 sources (34$\%$) are dominated by cirrus components,
mostly cool ($\psi$=1) or cold ($\psi$=0.1) dust (only 46 of these were fitted by the warmer ($\psi$=5) dust characteristic of our Galaxy).  
Thus 500$\mu$m selection favours galaxies with cooler interstellar dust than our own Galaxy.

By contrast, for 60303 {\it IRAS} Faint Source Catalog galaxies selected at 60$\mu$m (Wang and
Rowan-Robinson 2009, 2014), 8$\%$ are ultraluminous and 0.7$\%$ are hyperluminous.  Only
4 IRAS FSC galaxies are definitely known to be lensed.  42$\%$ are dominated by cirrus components, but only
6$\%$ of these cirrus galaxies require cooler dust.  Just 4$\%$ of IRAS galaxies have z $>$ 0.3, compared with 88$\%$ of unlensed 500$\mu$m galaxies.  
Thus, even allowing for the fact that the 500$\mu$m survey reaches a deeper galaxy density than the IRAS FSC 60$\mu$m survey, it is clear that 
a wavelength of 500$\mu$m provides a dramatically different picture of the infrared galaxy population to that seen by {\it IRAS} at 60$\mu$m.

\begin{figure*}
\includegraphics[width=7cm]{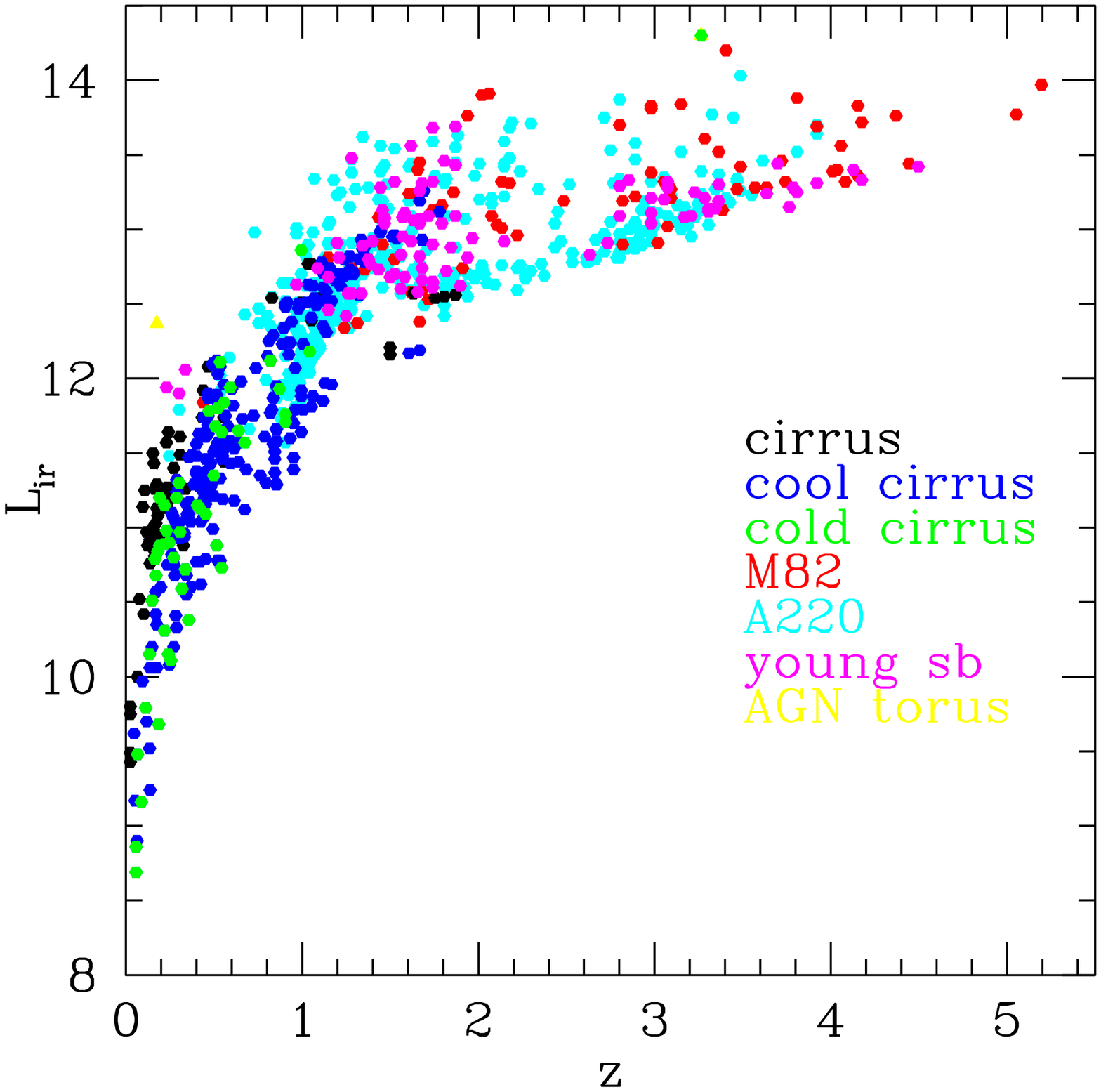}
\includegraphics[width=7cm]{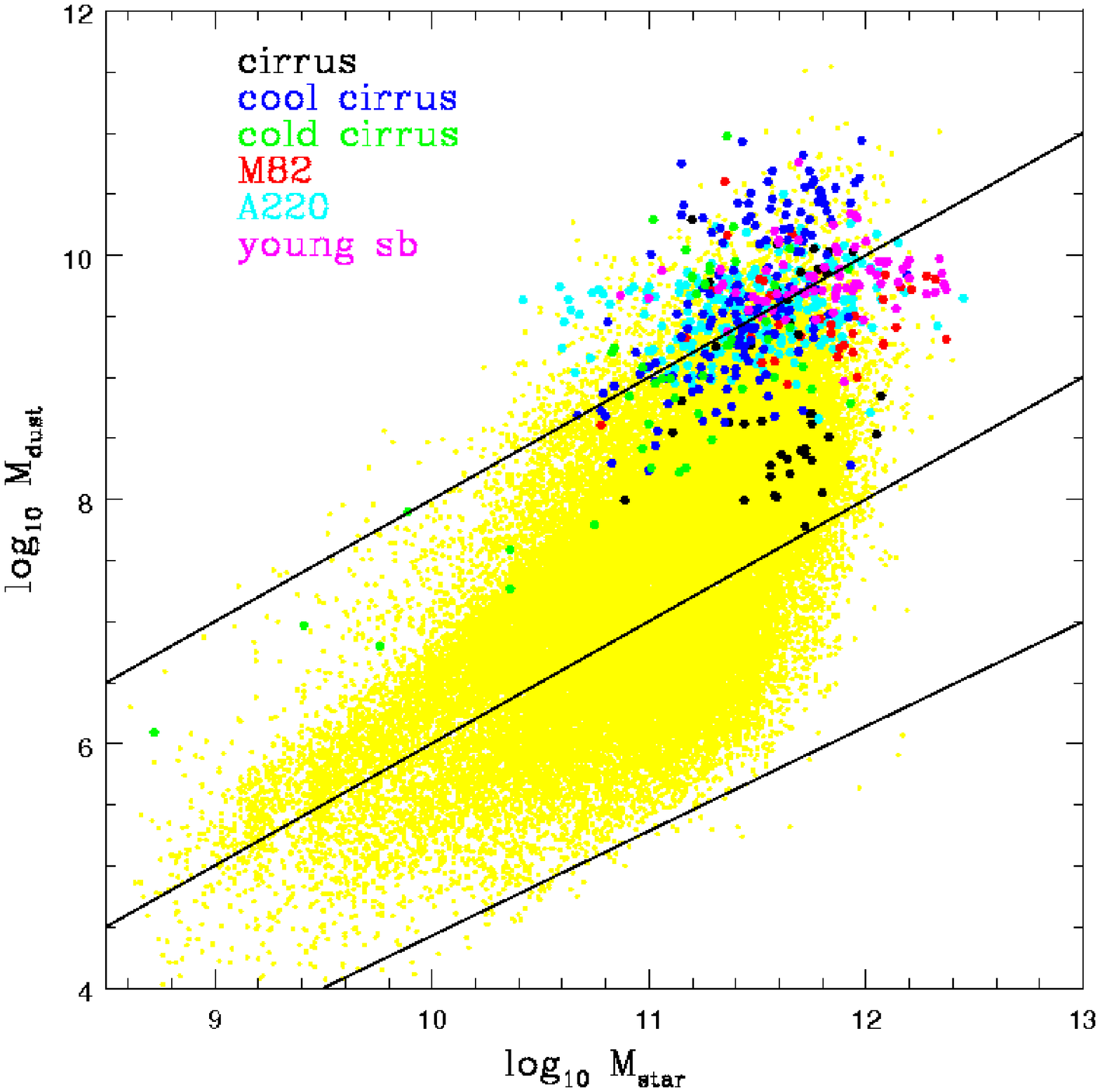}
\caption{LH: The distribution of infrared luminosity, $L_{ir}$, with redshift for the 858 SCAT 500$\mu$m 
sources in Lockman which are associated with unlensed SWIRE galaxies.  Black: cirrus ($\psi$=5); blue: cirrus ($\psi$=1); 
green: cirrus ($\psi$=0.1); red: M82
starburst; cyan: Arp220 starburst; magenta: young starburst; yellow: AGN dust torus.
RH: $log_{10} M_{\rm dust}$ v. $log_{10} M_{\rm star}$, in solar units, for HerMES-Lockman galaxies superposed on 
distribution for whole SWIRE catalogue (yellow points).
}
\end{figure*}

Having submillimetre fluxes not only gives us key diagnostic information on the nature of the infrared emission from galaxies, it
also allows us to estimate the dust mass far more accurately.  Fig 25R shows dust mass, calculated from
our radiative transfer models as in Rowan-Robinson et al (2010, 2013), versus stellar mass calculated from our stellar synthesis
optical-nir galaxy templates (Rowan-Robinson et al 2008).  The Herschel galaxies lie at the upper end of the dust-mass distribution
seen for all SWIRE galaxies, shown as a yellow distribution.  Much deeper submillimetre surveys would be needed to sample
the full range of dust-masses in galaxies.

Our results are not directly comparable with most other studies of the SEDs of SPIRE sources.
Hwang et al (2010), Symeonidis et al (2013) and Magnelli et al (2014) fit SEDs with a modified blackbody.  Conclusions about
correlation of 'dust temperature' with infrared luminosity or redshift correspond in our approach to variations in the proportions
of different components.  The modified blackbody approach gives no insight into whether the dust has high or low optical depth,
or whether the galaxy is in a starburst or quiescent phase.   Magdis et al (2013) have studied a sample of 330 galaxies with
S(24) $>$ 5 mJy in the SWIRE areas for which {\it Spitzer}-IRS data are available.  Just 2$\%$ of our sample are this bright at
24$\mu$m and there is no overlap with our sample. Their conclusions mainly concern PAH strength and AGN content.

We show the 500$\mu$m differential source-counts for 858 unlensed Lockman-SCAT-SWIRE sources + the 368 unassociated 
SPIRE sources, and for the 109 candidate
lensed galaxies (Fig 26).   The total number of potential lenses, 24 $\mu$m galaxies with redshift between 0.15 and 0.95
is 28549, so the total area surveyed for lenses is found by multiplying this by the solid angle subtended by a typical
Einstein ring.  For $z_{lens} \sim$ 0.5, $z_{image} \sim$ 3.0, $M_{lens} = 4.10^{11} M_{\odot}$ (allowing for the dark matter contribution), we estimate the radius of a
typical Einstein ring as $\sim$1.4 arc sec.  This yields a fraction of the 7.53 sq deg surveyed for lenses as 1:520.
This estimate is consistent with the lensing probability estimated from the CLASS survey by Browne et al (2003) of 1:690$\pm$190.
From Wardlow et al (2013) we estimate that a mean lensing magnification of 5 is appropriate for our S(500)=25mJy selection.
As a consistency check we have therefore shown the lensed
galaxy counts corrected by this amount in flux-density, and by the factor 520 in number, i.e. we are demagnifying the observed counts
of lensed objects and scaling by the lensed fraction to give a crude estimate of the unlensed counts for these objects.  
The resulting combined lensed and 
unlensed counts are plausible when compared with a compilation of observed counts and with source-count models (eg Rowan-Robinson 2009). However the are also consistent with as many as 50$\%$ of lens candidates with S500 $<$ 50 mJy being chance associations. 

Gonzalez-Nuevo et al (2012) have used SED arguments (but without the benefit of the all-important 3.6-24$\mu$m data)
to define lens candidates down to S500 $\sim$ 60 mJy in the H-ATLAS survey, at a surface density of $\sim$ 2 sources per sq deg.  
Our lens candidate sample is much deeper, at 25 mJy, and with a surface-density  $\sim$14 per sq deg.

\begin{figure*}
\includegraphics[width=14cm]{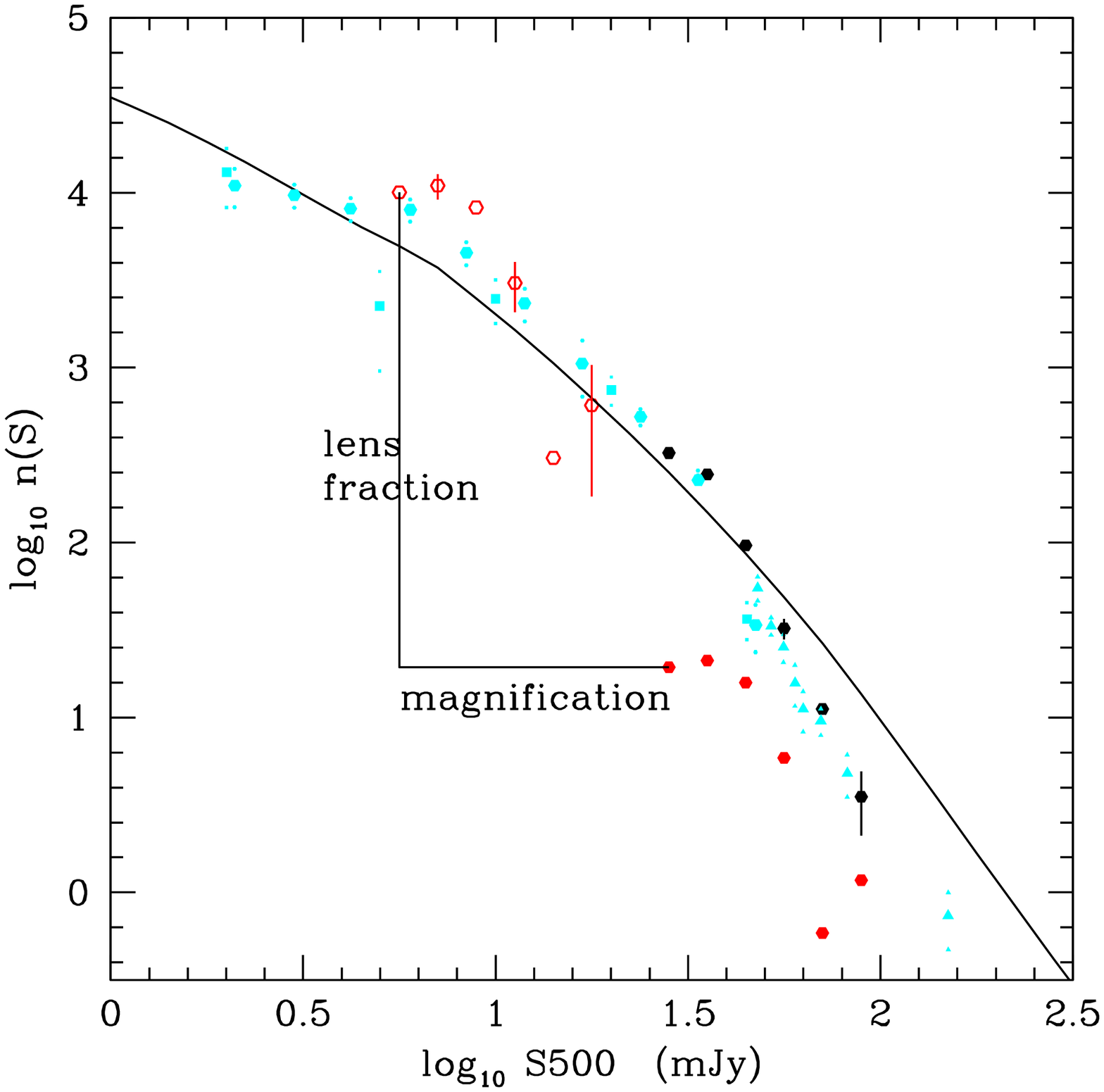}
\caption{The 500$\mu$m differential counts (n(S)=dN$/$lnS, sq deg$^{-1}$)for 858 unlensed Lockman-SCAT-SWIRE sources
+ the 368 unassociated sources (black filled circles) and lensed galaxies (red filled
circles),  The latter have also been shown corrected for an average lensing magnification of 5, and with the numbers multiplied
by 520 to correct for the fraction of the Lockman area that is imaged by potential lensing galaxies (red open circles).  
A compilation of HerMES and ATLAS differential counts (Bethermin, private communication) is also shown (cyan symbols).
A representative count model prediction (Rowan-Robinson 2009) is also shown (solid curve).
}
\end{figure*}

\section{Conclusion}
We have studied in detail a sample of 967 SPIRE sources with 5-$\sigma$ detections at 350 and 500$\mu$m, and associations
with SPIRE 24$\mu$m galaxies, in the HerMES-Lockman survey area, fitting their mid and far infrared, and submillimetre, SEDs
with a set of seven infrared templates.  For almost 300 galaxies we  have modelled their SEDs individually.  We confirm the
need for the new cool and cold cirrus templates, and also of the young starburst template, introduced by Rowan-Robinson et al 
(2010).  We also identify 109 lensing candidates via their anomalous SEDs and provide a set of colour-redshift constraints
which allow lensing candidates to be identified from combined {\it Herschel} and {\it Spitzer} data.  The lensing candidates and the 
galaxies requiring cold dust to understand their SEDs need to be confirmed with submillimetre interferometry, optical or near infrared imaging, and submillimetre or optical spectroscopy.

The picture that emerges of the submillimetre galaxy population is complex, comprising ultraluminous and hyperluminous starbursts,
lower luminosity galaxies dominated by interstellar dust emission, lensed galaxies and galaxies with surprisingly cold (10-13K) dust.
11$\%$ of 500$\mu$m selected sources are lensing candidates.  70$\%$ of the unlensed sources are ultraluminous infrared
galaxies and 26$\%$ are hyperluminous.  34$\%$ are dominated by optically thin interstellar dust ('cirrus') emission, but
most  of these are due to cooler dust than is characteristic of our Galaxy.   At the highest infrared luminosities we see SEDs
dominated by M82, Arp220 and young starburst types, in roughly equal proportions.

\section{Acknowledgements}
SPIRE has been developed by a consortium of institutes led by
Cardiff Univ. (UK) and including Univ. Lethbridge (Canada);
NAOC (China); CEA, LAM (France); IFSI, Univ. Padua (Italy);
IAC (Spain); Stockholm Observatory (Sweden); Imperial College
London, RAL, UCL-MSSL, UKATC, Univ. Sussex (UK); Caltech, JPL,
NHSC, Univ. Colorado (USA). This development has been supported
by national funding agencies: CSA (Canada); NAOC (China); CEA,
CNES, CNRS (France); ASI (Italy); MCINN (Spain); SNSB (Sweden);
STFC (UK); and NASA (USA).  The Dark Cosmology Centre (JW) is
funded by the Danish National Research Foundation.

The data presented in this paper are available through the {\it Herschel}  
Database in Marseille HeDaM\footnote{hedam.oamp.fr/HerMES}.

SJO acknowledges support from the Science and Technology Facilities Council (grant
numbers ST/L000652/1) and from the European Commission Research Executive Agency REAÊ
(Grant Agreement Number 607254).

E. Ibar acknowledges funding from CONICYT/FONDECYT postdoctoral project
N$^\circ$:3130504.


\end{document}